%% file: paper.tex
\documentclass[sigconf]{acmart}
\usepackage{listings}
\usepackage[mathscr]{euscript}
\usepackage{amsfonts}
\usepackage{amsthm}

\usepackage{amssymb}
\usepackage{xfrac}

\newcommand{\code}[1]{\mbox{\texttt{#1}}}
\newcommand{\DI}{\Delta_I}

\def\ojoin{\setbox0=\hbox{$\bowtie$}%
  \rule[-.02ex]{.25em}{.4pt}\llap{\rule[\ht0]{.25em}{.4pt}}}
\def\leftouterjoin{\mathbin{\ojoin\mkern-5.8mu\bowtie}}

\newtheorem{theorem}{Theorem}

\copyrightyear{2025}
\acmYear{2025}
\setcopyright{cc}
\setcctype{by-nc-sa}
\acmConference[SIGMOD-Companion '25]{Companion of the 2025 International Conference on Management of Data}{June 22--27, 2025}{Berlin, Germany}
\acmBooktitle{Companion of the 2025 International Conference on Management of Data (SIGMOD-Companion '25), June 22--27, 2025, Berlin, Germany}
\acmDOI{10.1145/3722212.3724455}
\acmISBN{979-8-4007-1564-8/2025/06}

\begin{document}

\title[Streaming Democratized with Dynamic Tables]{Streaming Democratized: Ease Across the Latency Spectrum with Delayed View Semantics and Snowflake Dynamic Tables}


\author{Daniel Sotolongo        }  
\authornote{Coauthored this paper.}
\author{Daniel Mills         } 
\authornotemark[1]
\author{Tyler Akidau         } 
\authornotemark[1]

\author{Anirudh Santhiar     } 
\authornotemark[1] 
\author{Attila-Péter Tóth    }
\authornotemark[1]
\author{Ilaria Battiston     } 
\authornotemark[1]
\authornote{Part of CWI; work conducted at Snowflake.}
\author{Ankur Sharma         } 
\author{Botong Huang         } 
\author{Boyuan Zhang         }
\affiliation{%
  \institution{Snowflake}
}

\author{Dzmitry Pauliukevich }
\author{Enrico Sartorello    }
\author{Igor Belianski       }
\author{Ivan Kalev           }
\author{Lawrence Benson      } 
\author{Leon Papke           } 
\author{Ling Geng            }
\author{Matt Uhlar           } 
\author{Nikhil Shah          }
\author{Niklas Semmler       } 
\affiliation{%
  \institution{Snowflake}
}

\author{Olivia Zhou          }
\author{Saras Nowak          }
\author{Sasha Lionheart      } 
\author{Till Merker          }
\author{Vlad Lifliand        }
\author{Wendy Grus        }
\author{Yi Huang             }
\author{Yiwen Zhu            }
\affiliation{%
  \institution{Snowflake}
}
\email{dynamic-tables-sigmod-DL@snowflake.com}

\renewcommand{\shortauthors}{Sotolongo, Mills, Akidau, et al.}

\begin{abstract} 
Streaming data pipelines remain challenging and expensive to build and maintain, despite significant advancements in stronger consistency, event time semantics, and SQL support over the last decade. Persistent obstacles continue to hinder usability, such as the need for manual incrementalization, 
semantic discrepancies across SQL implementations, and the lack of enterprise-grade operational features (e.g. granular access control, disaster recovery). 
While the rise of incremental view maintenance (IVM) as a way to integrate streaming with databases has been a huge step forward, transaction isolation in the presence of IVM remains underspecified, which leaves the maintenance of application-level invariants as a painful exercise for the user. 
Meanwhile, most streaming systems optimize for latencies of 100 milliseconds to 3 seconds, whereas many practical use cases are well-served by latencies ranging from seconds to tens of minutes.

In this paper, we present delayed view semantics (DVS), a conceptual foundation that bridges the semantic gap between streaming and databases, and introduce Dynamic Tables, Snowflake’s declarative streaming transformation primitive designed to democratize analytical stream processing. DVS formalizes the intuition that stream processing is primarily a technique to eagerly compute derived results asynchronously, while also addressing the need to reason about the resulting system end to end. Dynamic Tables then offer two key advantages: ease of use through DVS, enterprise-grade features, and simplicity; as well as scalable cost efficiency via IVM with an architecture designed for diverse latency requirements.

We first develop extensions to transaction isolation that permit the preservation of invariants in streaming applications.
We then detail the implementation challenges of Dynamic Tables and our experience operating it at scale. 
Finally, we share insights into user adoption and discuss our vision for the future of stream processing.

\end{abstract}



\received{9 December 2024}
\received[revised]{9 April 2025}
\received[accepted]{13 April 2025}

\maketitle

\section{Introduction}

As modern applications evolve, they increasingly rely on the continuous
transformation of data to generate insights, optimize operations, and enhance
user experiences. Stream processors play a pivotal role in enabling these
transformations, particularly for workloads requiring latencies below a few
tens of minutes. However, building and maintaining streaming pipelines remains
a significant technical challenge.

In practice, stream processors are often reserved for high-value use cases that
justify an investment in specialized engineering teams. These teams must
navigate the complexities of frameworks designed specifically for stream
processing, shouldering the cognitive overhead of their nuanced semantics. Most
streaming systems are stateful, long-lived, and optimized for sub-second
latencies, creating an impedance mismatch for use cases with more relaxed
latency requirements. Consistency guarantees are almost always 
framed as a choice between ``at-least once'' and ``exactly-once'', with no concept of transaction isolation, and the preservation of application invariants delegated to users. 
Furthermore, these systems often lack enterprise-grade
features such as centralized data catalogs, fine-grained access
control, and built-in disaster recovery. As a
result, streaming transformations are underrepresented, even in scenarios where
they could provide substantial value.

\textbf{Dynamic Tables} (DTs) is Snowflake's solution to these challenges, simplifying
stream processing to the level of writing a SQL query. The
architecture of DTs is designed for processing latencies from
seconds to hours, leveraging the cloud’s elasticity to facilitate resource
reuse. Workloads are fully orchestrated by Snowflake, guided by
the user-specified target lag. Additionally, DTs integrates
seamlessly with Snowflake's enterprise-grade features, facilitating adoption.

Instrumental to the simplicity of Dynamic Tables
is \textbf{delayed-view semantics}, a conceptual foundation that bridges the gap 
between traditional database semantics and stream processing, and which DTs
implements cost-effectively through automatic incremental view maintenance (IVM). With delayed view semantics, the contents of a DT are guaranteed to be logically equivalent to a corresponding view \textit{at some point in the past}. The simplicity of this guarantee facilitates reasoning about application invariants. On this foundation, we extend the robust body of transaction isolation research to cleanly model and explain the transactional phenomena that can arise in the context of stream processing in an RDBMS. 

The rapid adoption of Dynamic Tables underscores the accessibility of our
solution and the simplicity afforded by delayed view semantics. In the nine months following general availability, thousands of
Snowflake customers adopted DTs, with over one million active
tables. Customers consistently cite ease of use as the primary driver of
adoption. By cleanly embedding stream processing directly within the analytical DBMS,
DTs democratizes the paradigm, opening the door for a broader range
of practitioners to apply stream processing.

In this paper, we present the design of Dynamic Tables and delayed view semantics, as well as the rationale behind their key concepts (\S\ref{sec:concepts}). We look at delayed view
semantics in detail, and adapt the traditional model of transaction isolation to it (\S\ref{sec:iso}).
We then detail the architecture and implementation of
DTs, highlighting the challenges overcome and opportunities for
future research (\S\ref{sec:implementation}). Finally, we share insights from
the development, launch, and operation of DTs, including how we ensure
system safety and liveness, an analysis of user adoption, and an evaluation of Incremental View Maintenance as a theoretical foundation for stream processing.

\section{Related Work} \label{sec:related-work}
Materialized Views (MVs) and Incremental View Maintenance (IVM) have been extensively studied in both academia and industry. Traditionally, MVs were seen only as a way to accelerate interactive queries, and IVM was restricted to simple operations. Research between 1980 and 2010~\cite{shmueli84maintenance, blakely86efficient, griffin95dups, gupta99maintenance, Zhou2007LazyMO} laid a foundation for modern IVM frameworks, establishing the problem, proposing algorithms for efficient maintenance, and moving processing out of the critical update path.

Over the last decade, research into IVM has broadened its scope. In 2012, DBToaster~\cite{dbtoaster} introduced the concept of higher-order IVM to dramatically accelerate the maintenance of join-aggregate queries. In 2013, Differential Dataflow~\cite{mcsherry2013differential} introduced a novel approach to IVM for recursive queries. In 2018, Noria~\cite{gjengset2018noria} invented partial IVM in support of request-serving workloads. In 2023, DBSP~\cite{budiu2022dbsp} introduced an elegant, general framework to incrementalize computations in terms of commutative groups.

Meanwhile, large database vendors have long supported MVs in some form: Oracle~\cite{bello98mvoracle}, Microsoft SQL Server~\cite{site-microsoft-mv}, AWS Redshift~\cite{site-redshift-mv}, Google BigQuery~\cite{site-google-mv}, and more. These implementations are focused on the traditional use case of accelerating interactive queries, and are missing features necessary for use in data pipelines, such as nesting, comprehensive query coverage, and control over resources.
The last half-decade has birthed a cornucopia of IVM-based stream processing systems, led by torchbearer Materialize~\cite{mcsherry2022materialize} (based on Differential Dataflow), and now comprising a crowded startup market including Rising\-Wave, SingData, TimePlus, and Feldera (based on DBSP), most of which are focused on sub-10-second latencies.
In 2022, Databricks announced the general availability of Delta Live Tables~\cite{site-dlt}, which is an MV-centric, data pipeline product which targets the same set of use cases as Dynamic Tables, but has nonstandard SQL semantics. 

Until recently, the stream processing community was largely separate from the database community, with a history of programming-language-embedded frameworks such as MillWheel~\cite{akidau2013millwheel}, Storm~\cite{toshniwal2014storm}, Kafka~\cite{kreps2011kafka}, Flink~\cite{carbone2015apache}, Google Cloud Dataflow~\cite{akidau2015dataflow}, and Spark Structured Streaming~\cite{armbrust2018structured}. These systems arose from the Big Data and NoSQL movements, and were primarily concerned with achieving horizontal scalability. Each advanced the state of the art of stream processing, but proposed distinct APIs and semantics which created barriers for use in practice. In the last several years, these systems have converged on SQL as their primary interface, and invested heavily in improving usability.

Dynamic Tables is an implementation of the abstract concept of an MV. It surfaces stream processing capabilities directly within the Snowflake analytical RDBMS~\cite{snowflake}, implementing standard SQL semantics by repurposing Snowflake's existing capabilities and extending them~\cite{akidau2023difference} to include IVM and orchestration of complex pipelines. Delayed View Semantics permits the precise specification of the transaction isolation characteristics of materialized views that provide out-of-date results, such as DTs.

\section{Concepts} \label{sec:concepts}
In this section, we describe the semantics of Dynamic Tables in terms of user-visible concepts, detailing what, when, and how they compute.

A DT is represented as a table in the Snowflake RDBMS, and its contents are the result of its defining query at some point in the past. To create it, a user provides a \code{SELECT} query, a target lag duration, and a virtual warehouse in which to execute refreshes. Once created, Snowflake can \emph{refresh} the DT, computing the result of the provided query, and storing its results in the DT. A refresh can be triggered automatically to meet the \emph{target lag}, or by the user via a \emph{manual refresh}. For a large class of queries, Snowflake performs an \emph{incremental} refresh, computing the changes since the last refresh and applying those changes to the already-stored result. For other queries, Snowflake performs a \emph{full} refresh by executing the query from scratch. A DT can be queried like any other table, and it provides fast, predictable performance because its contents have been precomputed. Listing~\ref{lst:dt} shows an example of creating a pair of DTs that keep track of late-arriving trains. Once created, these DTs will automatically, incrementally refresh to keep their contents no more than 1 minute out of date.

\begin{lstlisting}[caption={Example Dynamic Table definitions.}, label={lst:dt}]
CREATE DYNAMIC TABLE train_arrivals
  TARGET_LAG = DOWNSTREAM
  WARHEOUSE = trains_wh
  AS SELECT
    t.id train_id,
    e.payload:time::timestamp arrival_time,
    e.payload:schedule_id::int schedule_id
  FROM train_events e
  JOIN trains t ON e.payload:train_id::int = t.id
  WHERE e.type = 'ARRIVAL';
  
CREATE DYNAMIC TABLE delayed_trains
  TARGET_LAG = '1 minute'
  WAREHOUSE = trains_wh
  AS SELECT train_id, 
      date_trunc(hour, s.expected_arrival_time) hour,
      count_if(arrival_time - s.expected_arrival_time > '10 minutes') num_delays
    FROM train_arrivals a
    JOIN schedule s ON a.schedule_id = s.id
    GROUP BY ALL;
\end{lstlisting}

\subsection{\textit{What} is computed?}
When first created, a DT is \emph{initialized} with the result of its defining query. Initialization can be done either synchronously, as part of the creation, or asynchronously according to the target lag. Querying a DT before it has been initialized results in an error.

\subsubsection{Delayed View Semantics (DVS)}
After a DT is initialized, its contents are guaranteed to be the result of its defining query at some point in the past. This time is called the DT's \emph{data timestamp}. We call this guarantee \emph{delayed view semantics}, which communicates the idea that a DT is logically equivalent to a view, but with results delayed by some duration. Whenever the DT is refreshed, its data timestamp is updated.

The defining query of a DT can read base tables, views, and other DTs. We say that sources are \emph{upstream} of a DT, and that a DT is \emph{downstream} of its sources. Cycles are not allowed. Reading from other DTs in a DT's defining query creates an ambiguity in the definition of delayed view semantics: when it refreshes, what are the contents of the DTs it reads from? Two possibilities seem natural. The first, which we call \emph{persisted table semantics}, simply reads the data that was persisted in the upstream DT at the time of the downstream's refresh. The second, which we propose as the definition of \textit{delayed view semantics}, reads the data in the DT corresponding to the same data timestamp as the current refresh. See \S\ref{sec:iso} for more about this distinction. Dynamic Tables implements DVS.

\subsubsection{Data Timestamp Selection}
Read dependencies between DTs induce a directed acyclic graph, where tables, views, and DTs are vertices, and edges represent dataflow between them. In order to ensure delayed view semantics, Snowflake resolves this graph as of a given data timestamp and refreshes each DT after all of its upstream dependencies have data available at that timestamp. This applies to initializations, scheduled refreshes, and manual refreshes. 

Manual refreshes choose a data timestamp that is after the refresh command was issued. This generally requires a refresh to be run for all DTs upstream of the one being refreshed.

Scheduled refreshes grant Snowflake the flexibility to choose the data timestamp, as long as the timestamp chosen for each DT is also chosen for those upstream. We discuss how we make this choice in \S\ref{sec:sched}.

Initializations present a challenge. An initialization could be treated as a regular manual refresh, choosing a data timestamp around the creation time and refreshing all upstream DTs as of that time. However, a very common pattern for creating DTs is to create them in dependency order. Referring to Listing \ref{lst:dt}, the typical user would create \code{train\_arrivals}, then \code{delayed\_trains}. Choosing a new timestamp for each initialization would refresh \code{train\_arrivals} twice for no reason, and the number of refreshes increases quadratically with the depth of the graph. Therefore, Snowflake chooses an initialization timestamp to minimize the amount of wasted computation: the most recent data timestamp of its upstream DTs that is within the target lag, or the creation time if none exists. This approach yields a result that is correct and meets the users' target lag, but has the counterintuitive consequence that a DT created at $t$ might be initialized to a data timestamp of $t' < t$. We have found this to be a small sacrifice for the clean semantics afforded by delayed view semantics.

\subsection{\textit{When} is it computed?}
Delayed view semantics defines the contents of a DT, but users also care about when results are available. This is controlled through the \emph{target lag} property, which defines how up-to-date the table’s contents should be. \emph{Lag} (also known as ``freshness'' or ``staleness''), is the difference between the current time and the table’s data timestamp. In other words, the lag is the delay in the content of the DT with respect to its base tables. The target lag instructs Snowflake to maintain the table’s lag below the given value.

Dynamic Tables support two types of target lags: a duration or \code{DOWNSTREAM}. Durations (minimum of 1 minute, support for lower values is in early testing) specify a time-based lag limit, subject to upstream table constraints. The \code{DOWNSTREAM} option automatically aligns the table’s lag with the minimum target lag of its downstream dependencies, simplifying configuration by making the table refresh only when required by others.

Lag has traditionally been used in stream processing as a system health indicator, with rising lag signaling degradation. Exposing lag as a user-defined parameter is less common. Instead, many systems expose tuning parameters like refresh periods or buffering intervals, which require manual calibration. Snowflake simplifies this by exposing target lag --- a user-intuitive concept --- as a direct configuration parameter. This aligns with Snowflake’s philosophy of eliminating implementation-dependent tuning, allowing users to specify intent clearly. This intent gives Snowflake the freedom to unlock significant cost savings, such as redirecting compute resources during low-activity periods.

While target lag supports diverse latency requirements, it doesn’t cover all use cases. Batch-oriented pipelines often operate on cron-based schedules, such as when data arrives in bulk at night. For example, a full-refresh table joining two datasets updated at 1:00 and 2:00 would waste resources if refreshed between these times. For such cases, Snowflake provides a manual refresh operation, configurable via Snowflake Tasks with cron-based schedules. Future plans aim to integrate this functionality directly into Dynamic Tables for greater simplicity.

\subsection{\textit{How} is it computed?}
While Dynamic Tables hides substantial complexity, users can still benefit from understanding the system's operational characteristics. These include how computational resources are managed, how the system responds under load, and how the cost of refreshes varies across different contexts.

\subsubsection{Warehouses}
Snowflake's architecture separates the multi-tenant control plane, which parses and optimizes queries, executes control commands, and stores metadata, from the single-tenant data plane, which executes optimized query plans. The control plane, which we call \emph{Cloud Services}, is completely outside of users' control, and Snowflake takes responsibility for its management. In contrast, responsibility for the data plane is shared with users. Snowflake provides a catalog entity called a \emph{Virtual Warehouse}, which represents a cluster of nodes that can execute queries. Snowflake charges for the time a virtual warehouse is active at a granularity of seconds. Virtual warehouses can be started, suspended, and resized on demand, and support automatic suspension when inactive. The choice of virtual warehouse does not limit access to data in the account. Each warehouse belongs to a single customer account and is owned by a single role, which controls access to its resources. Warehouses allow users to trade resource utilization and latency by isolating or co-locating their workloads.

When creating a DT, the user must provide a warehouse in which to execute its refreshes. This gives Snowflake customers a familiar interface, and provides the same capabilities for workload management as standard Snowflake queries. Common patterns include co-locating a set of related DTs in the same warehouse for cost efficiency and setting a low auto-suspend duration when a workload is expected to only be run intermittently.

\subsubsection{Refresh Modes}
Each DT has a \emph{refresh mode}, which can be either \code{FULL} or \code{INCREMENTAL}. Incremental mode is currently supported for projections, filters, union-all, inner and outer joins, \code{LATERAL FLATTEN}\;(for dealing with nested data), distinct and grouped aggregations, and partitioned window functions. It is not yet supported for scalar subqueries, \code{[NOT] (IN | EXISTS)}, scalar aggregates, or various specialized operators like \code{ASOF} joins, \code{MATCH\_}\allowbreak\code{RECOGNIZE}, etc. 
We do not expect users to understand IVM to any degree of depth. Instead, we communicate the scaling behavior of these refresh modes. Full refreshes behave in a straightforward way, with cost similar to computing the result of the defining query. The cost of incremental refreshes depends on many variables, but we can simplify it to fixed and variable costs. Generally, more complex queries have larger costs (both fixed and variable), and variable costs scale linearly with the amount of changed data in the sources. This basic idea is usually sufficient for customers to find successful configurations.

A DT refresh can take one of several \emph{actions}. If none of the data sources changed since the last refresh, it will perform a \code{NO\_DATA} refresh, which only updates the DT's data timestamp and does not consume any virtual warehouse resources. If the source data has changed and the refresh mode is \code{FULL}, the action will also be \code{FULL}, which is essentially an \code{INSERT OVERWRITE} using the defining query at the refresh's data timestamp. On the other hand, if the refresh mode is \code{INCREMENTAL}, the action may either be \code{INCREMENTAL} or \code{REINITIALIZE}. The former is the typical case, and works by computing the changes in the defining query since the last refresh's data timestamp, and then applying those changes to the DT. The latter is similar to a full refresh, and is used when some change upstream, such as replacing an upstream table, may have invalidated the results already stored in the DT, at which point an incremental refresh would cause data corruption.

\subsubsection{Skips and Errors}
When the resources required to timely refresh a DT exceed the resources allocated to it, refresh operations will exceed their allotted time. The current implementation of Dynamic Tables does not permit concurrent refreshes of the same DT, so subsequent refreshes cannot execute until the preceding one completes. When such a situation is detected, Snowflake chooses to \emph{skip} the later refresh, relying on the subsequent refresh to bring the DT's data timestamp up to date. Note that a skipped refresh does not compromise on delayed-view semantics. A refresh following a skip upholds the same guarantees by including the skipped time interval into its change interval. A skip does reduce the granularity of the DT's time travel history, which will not have an entry for the skipped data timestamp. For most incrementally-maintained DTs, a skip also means the next refresh will have to do more work. That said, skipping a refresh reduces the total amount of work by eliminating the fixed costs of the skipped refresh. This property allows DTs to gracefully increase their rate of progress as they fall further behind.

If a refresh encounters a user error, such as division-by-zero, it fails and is not retried. The next scheduled refresh (for a different data timestamp) will be attempted, with each consecutive failure incrementing an error counter for the DT. If the counter exceeds a threshold, the DT is automatically suspended to prevent wasting compute resources. If the root cause of the problem is in the underlying data or upstream metadata, the DT can resume from where it left off once the cause is addressed. If the problem is in the defining query itself, the owner of the DT can replace the DT. 

Responsibility over the operational health of a DT is shared between Snowflake and the user. Snowflake ensures that refreshes are scheduled at appropriate times, produce correct results, and have comparable performance to an equivalent query outside of the context of Dynamic Tables. Users must ensure that the target lag requirement is achievable given the combination of query, data, and resources, which is often achieved through experimentation. There are numerous opportunities for improvement in this area. Automatically selecting the warehouse size based on cost estimates is a clear opportunity, but estimating the cost of incremental plans is a challenging technical problem. Additionally, subtle details in the design of incrementalization can significantly affect the predictability and stability of refreshes. As a result, we now prioritize stable performance over peak performance when implementing new or better incrementalizations.

\subsection{Miscellaneous Features}
Dynamic Tables is just one construct in the broad set of features offered by Snowflake. Seamless integration with these other features is a crucial requirement for them to be easy to use.

Snowflake has extensive support for role-based access control (RBAC). In addition to \code{SELECT} and \code{OWNERSHIP}, DTs also provide \code{MONITOR} and \code{OPERATE} privileges, which allow grantees to see the current status of and invoke refreshes on a DT, respectively.

A challenging aspect of the design of SQL databases is the imperative nature of DDL, which means that a change to any entity in the DBMS may occur at any time. For Dynamic Tables, this challenge is particularly apparent when users alter, replace, or rename entities upstream of a DT. To address this problem, we adopted two principles. First, upstream dependencies take precedence over downstream. For example, if a centralized data engineering team needs to replace a table for some reason, they should not be prevented from doing so. Second, Dynamic Tables should recover automatically when such changes are made. For example, if a table is dropped, a DT refresh downstream of it will fail. But if the table is \code{UNDROP}ped, then refreshes should resume without issue. 

Snowflake supports zero-copy-cloning, whereby a new table, schema, or database is created with the contents of another by copying only its metadata. It also supports renaming, swapping, and cloning of databases and schemas, both of which can be spanned by DT graphs. When such an operation is performed, a whole subgraph of DTs is moved or created. Our implementation preserves delayed view semantics, continuing unperturbed if unaffected or reinitializing if the operation replaced any of their dependencies.  Cloned DTs can avoid reinitialization in many cases.

Snowflake offers several features made possible by its Cloud-native nature. Data sharing permits Snowflake users to share data across accounts. Dynamic Tables can be shared just like any other table or view. Cross-region replication of DTs allows users to easily move data between regions for sharing or disaster recovery, creating an unprecedented level of simplicity for global, highly available data platforms. Iceberg tables allow users to seamlessly query Snowflake tables from external query engines and query tables in other systems from Snowflake. Dynamic Tables can read from Iceberg Tables and can be stored as Iceberg Tables. 
Snowflake's goal for Iceberg is to provide performance parity with native tables, and is pursuing numerous proposals to bring proprietary features to the Iceberg Specification~\cite{iceberg-row-lineage}\cite{iceberg-variant}.

Snowflake supports running code in Python, Java, and Scala via Snowpark~\cite{snowpark}, which provides convenient APIs for creating user-defined functions~(UDFs), stored procedures, and data-frame-based queries. Dynamic tables can be defined using Snowpark APIs, and they can execute Snowpark UDFs. This enables, for example, the use of DTs to compute embeddings and run inference for AI/ML use cases.

UDFs raise the question of how non-determinism interacts with delayed view semantics. We have found that there are different kinds of nondeterminism, and users have different expectations for each. First, floating-point operations are nondeterministic when hardware and order are not fixed. Users rarely care about this, so we prohibit their use only when the nondeterminism would interfere with view maintenance, such as joining on a float aggregate key.
Second, context functions like \code{CURRENT\_TIMESTAMP} and \code{CURRENT\_ROLE} are not deterministic, but they do exhibit predictable behavior that users often expect Dynamic Tables to handle. We approach context functions on a case-by-case basis. Third, truly nondeterministic operations, such as UDFs that make remote calls or call random number generators, are usually expected to be run only when a row is inserted or updated in a DT. Dynamic tables do not yet support incremental refreshes in this case, but we expect to support it soon. Snowpark UDFs can be annotated as \code{IMMUTABLE} to indicate that they are deterministic, which enables incremental refreshes.

\section{Delayed View Semantics \& Transaction Isolation}
\label{sec:iso}
Delayed view semantics fits naturally within the transactional model implemented by RDBMSes, and Dynamic Tables guarantees ACID properties. So far, we have only described this semantics informally. In this section, we rigorously define how delayed view semantics fits with transaction isolation. We begin by describing in more detail the problem with persisted table semantics. We then propose our theoretical extensions that address the problem. Last, we use these extensions to describe precisely the semantics of Dynamic Tables.

\begin{figure}[h]
  \centering
  \includegraphics[width=\linewidth]{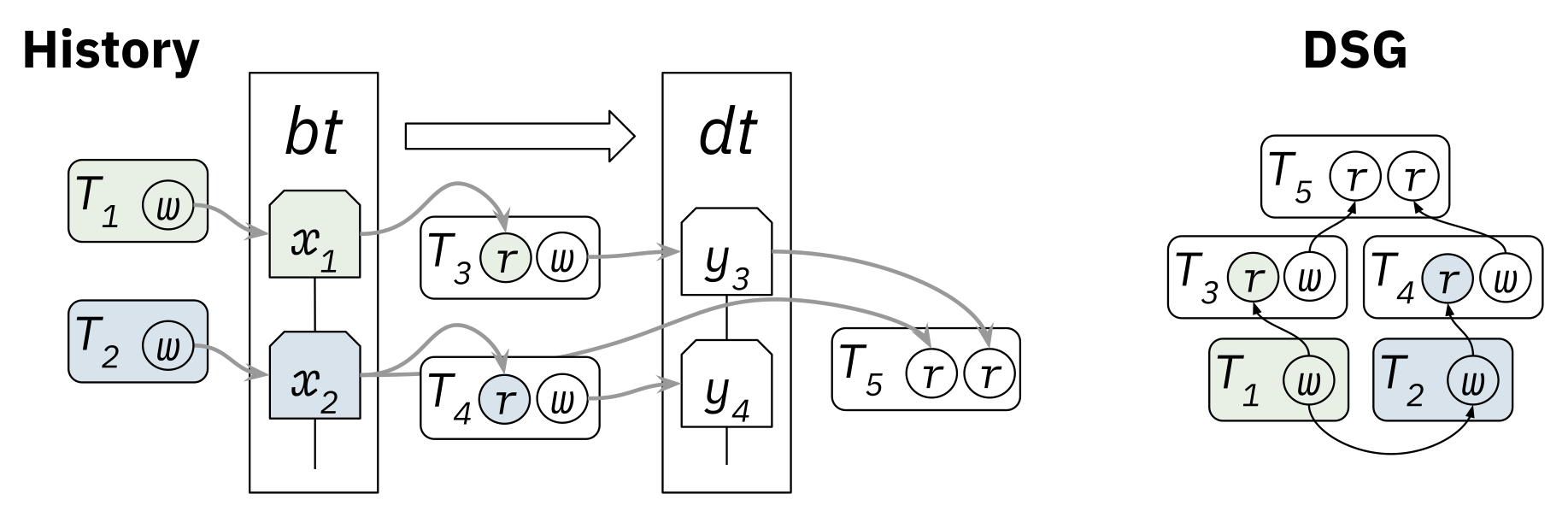}
  \caption{A diagram of persisted table semantics. The DSG is serializable despite the clear presence of read skew because the refresh transactions mask the conflict. }
  \label{fig:pts}
\end{figure}

The problem with persisted table semantics manifests because a DT refresh implementing it acts as a regular database transaction. This is simple and fits within existing literature on transaction isolation without modification: the isolation level of such a DT is simply the isolation level of the transactions enclosing its refreshes.
Unfortunately, from the perspective of the application, many unwanted phenomena can appear in the results of such DTs.
For example, consider the history in Figure \ref{fig:pts}. A dynamic table $dt$ reads from a base table $bt$, which has 2 versions of object $x$. 
These are written by transactions $T_1$ and $T_2$, respectively. $dt$ performs two refreshes to produce objects $y_3$ and $y_4$.
Then, another transaction $T_5$ reads $y_3$ and $x_2$. 
From the perspective of the application, $T_5$  observes read skew.
But the Direct Serialization Graph (DSG) shown on the right of the figure reveals that this history is, in fact, serializable.
The framework is unable to identify a phenomenon that seems obvious to observers.
Why?

Transactions were originally intended to model interactions between the database and its environment. Such interactions represent entirely new information flowing into the database, and the role of a transaction system is to prevent the database from accepting contradictory information. As a result of this framing, the traditional model assumes that every transaction, given a consistent database state, produces another consistent state. But the purpose of delayed view semantics is to move computation off of the critical path (for both writes and reads), into an intermediary transaction that, strictly speaking, creates an inconsistent state. The traditional model definitionally treats this inconsistent state as consistent, decoupling application-level and database-level notions of consistency. To restore this coupling, we extend the traditional transactional model to make a distinction between operations that interact with the environment and operations that are pure computation. The former must leave the database in a consistent state, but not the latter.

Intuitively, our extended transactional model merely allows us to reproduce delayed view semantics: querying a DT manifests the same transactional phenomena as querying an equivalent view whose results were delayed by the current lag. One might hope that we can simply pretend that the query actually was on a view, not a DT. But then the transaction history would not match the actual events that took place, which is unacceptable. So, our model introduces the concept of \emph{derived values}, which are values computed purely from data already stored in the DBMS, such as the contents of DTs. We define a new kind of operation, \emph{derivation}, which creates derived values and explicitly represents their provenance in the transaction history. Derivations allow us to trace dependencies between reads and writes that traverse derived values, mirroring the dependencies we would have observed if querying a delayed view. We can then identify isolation phenomena and speak in terms of isolation levels that are meaningful to applications, even when derived values are computed in separate transactions, as with DT refreshes.

The framework of Adya, et al.~\cite{adya} is the foundation for our extensions.
Adya defines the transaction history of a database as a partial order of events, which represent operations on objects by transactions, and a total order on the committed versions of each object. There are four kinds of operations: read, write, commit, and abort. Each event belongs to a single transaction. Using these events, they define three different kinds of dependencies between transactions, which generate a Direct Serialization Graph (DSG) from a history. Numerous phenomena are then defined in terms of the history or the DSG. Isolation levels are defined by proscribing specific phenomena from the possible histories of a database.

We extend Adya's model by adding a new kind of operation, \emph{derivation}, denoted: 
\[d_i(x_i | y_j^0, \dots, y_k^n)\]
This represents that the version $i$ of some object $x$ is a derived value, computed from versions $j \dots k$ of objects $y^0 \dots y^n$ in transaction $T_i$. We say an object $v_i$ \emph{derives from} another object $z_m$ when there exists a path of derivations connecting them:
\[d_i(v_i|x_j, \dots), d_j(x_j|\dots), \dots, d_k(y_k|z_m, \dots)\] 
We extend the definitions of the three kinds of dependency to include derivations:
\begin{itemize}
    \item We say that $T_j$ \emph{directly item-read-depends} on $T_i$ if $T_i$ installs some object version
$x_i$ and $T_j$ reads $x_i$ (prior definition), \textbf{or} if $T_i$ installs $y_k$, $T_j$ reads $x_i$, and $x_i$ derives from $y_k$.
    \item  We say that $T_j$ \emph{directly item-anti-depends} on $T_i$ if $T_i$ reads some object version $x_k$ and $T_j$ installs $x$’s next version (after $x_k$) in the version order (prior definition), \textbf{or} if $T_i$ reads some object version $x_k$, $x_k$ derives from an object version $y_m$, and $T_j$ installs $y$'s next version (after $y_m$).
    Note that the transaction that wrote the later version directly item-anti-depends on the transaction that read the earlier version.
    \item  A transaction $T_j$ \emph{directly write-depends} on $T_i$ if $T_i$ installs a version $x_i$ and $T_j$ installs $x$’s next version (after $x_i$ ) in the version order (prior definition), \textbf{or} if $T_i$ installs $x_i$, $T_j$ installs $y_j$, and there exist consecutive versions $z_k \ll z_m$ such that $z_k$ derives from $x_i$ and $z_m$ derives from $y_j$.
\end{itemize}
Definitions for predicate dependencies are similar. These definitions bring us to the key properties of derivations.

\begin{theorem}[Transaction Invariance]
    Given any history $H$ containing a transaction $T_i$ and a derivation $r = d_i(x_i | \dots)$, define another history $H'$ which moves $r$ into another transaction $T_j$ to create $d_j(x_j | \dots)$ and replaces all reads from $x_i$ in $H$ with reads from $x_j$. $H$ has exactly the same dependencies as $H'$.
\end{theorem}

This follows from the fact that the definitions of dependencies are agnostic to which transaction contains the derivation operation. Derivations represent a pure computation. They act as intermediaries, connecting transactions that perform reads with the transactions that wrote those values. Where that computation was done is irrelevant to that connection.

\begin{corollary}[Encapsulation]
    Given a history $H$ containing a transaction $T_i$ containing a derivation $d_i(x_i| y^n_j, ...)$, we say that $d_i$ is  \emph{encapsulated} by $T_i$ if it only reads values written by $T_i$ and its value is only read by operations in $T_i$. More precisely, if there exists no transaction $T_k$ in $H$, with $k \ne i$, which contains any of $r_k(x_i)$ or $w_k(y^n_j)$ for all corresponding $n$ and $j$.
    
    Every history $H'$ excluding an encapsulated derivation from a history $H$ has exactly the same dependencies as $H$. 
\end{corollary}

This corollary tells us that encapsulated derivations do not affect dependencies. Amusingly, this corollary allows us to pretend that derivations have been implicit in transactions all along, but always encapsulated. It was the need to un-encapsulate them by moving them to a separate transaction that forced us to acknowledge their existence!

The definitions of phenomena generalize nicely to include derivations. Here are the phenomena in \cite{adya}, updated to account for derivations. For all but G1b, the actual definitions are the same, but the presence of derivations in a history can induce new instances of the phenomena.

\begin{itemize}
    \item \textbf{G0: Write Cycle} occurs when the DSG contains a cycle of write dependencies. 
    \item \textbf{G1a: Aborted Read} occurs when a committed transaction read-depends on an aborted transaction.
    \item \textbf{G1b: Intermediate Read} occurs when a committed transaction reads an object version that is not the final version installed by a transaction (prior definition), \textbf{or} it reads an object that derives from such an intermediate version.
    \item \textbf{G1c: Circular Information Flow} occurs when the DSG contains a cycle of only read- and write-dependencies. These dependencies can be generated by paths of derivations.
    \item \textbf{G2: Anti-dependency Cycle} occurs when the DSG contains a cycle of read-, write-, and anti-dependencies.
\end{itemize}

\begin{figure}[h]
  \centering
  \includegraphics[width=\linewidth]{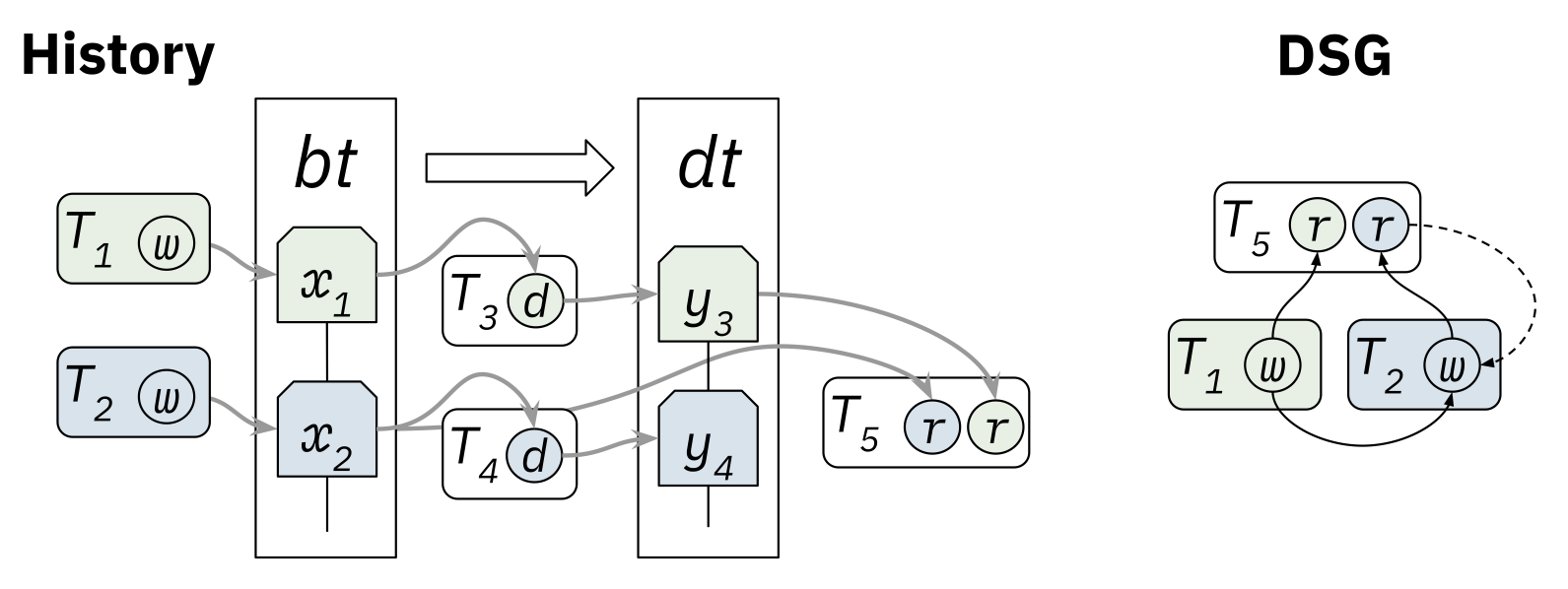}
  \caption{The example in Figure \ref{fig:pts}, mapped to delayed view semantics. The history now contains derivations, and the DSG contains a cycle, indicating read skew. }
  \label{fig:dvs}
\end{figure}

The example in Figure \ref{fig:dvs} reinterprets the example in Figure \ref{fig:pts} in terms of delayed view semantics. Instead of representing refresh operations as reads and writes, they are represented as derivations. This removes the refresh transactions from the DSG and generates an anti-dependency between $T_5$ and $T_2$, because $T_5$ read an object ($y_3$) which derives from an object ($x_1$) that was overwritten by $T_2$. This causes a cycle to appear, exhibiting phenomenon G2 (and G-single~\cite{adya1999weak}), and revealing the read skew that we knew was there all along.

This example demonstrates how using derivations to model asynchronous streaming computation empowers DBMS users to reason about application invariants and ensure consistency guarantees as developed in Adya's thesis~\cite{adya1999weak}. For instance, we expect that PL-2+ provides basic-consistency, even if histories contain derivations. We leave a full proof for future work.

In Snowflake, all DT refreshes consist exclusively of derivation operations, and no other transactions contain derivations. However, it would be feasible and logical to colocate refreshes in the same transaction as other DML operations.

Dynamic Tables provides two isolation levels in different contexts. If a transaction reads from a single DT (even if other DTs are upstream) and no other table, that transaction is guaranteed to have Snapshot Isolation (PL-SI). Otherwise, it is guaranteed Read Committed (PL-2). The reason for this distinction is that, if a transaction queries multiple DTs, the available data timestamps of the queried DTs may not match up. For example, suppose a query joins two DTs, one with a target lag of 1~minute and the other 1~hour. Should the query automatically rewind to the most recent common data timestamp? What if that timestamp is more than an hour in the past? Perhaps it should refresh one of the DTs to bring it up to date? Our implementation simply reads the current data, weakening the isolation level. We intend to further explore answers to this question in the future. For now, the current guarantee addresses a large set of use cases, since SI can be ensured by including a query of interest in a DT, and querying that directly.

\section{Implementation} \label{sec:implementation}
Dynamic Tables is designed to feel natural inside a SQL RDBMS and encapsulate the complexity of a streaming pipeline, while maximizing reuse of existing DBMS functionality. These principles strongly constrain the architecture and implementation of the system. Feeling natural in SQL means that a DT should look and feel like a table, that the semantics of its SQL query should closely match that of SQL queries in other contexts, and that any other features which make sense on a table should also be supported.
Encapsulating complexity means that all configuration should be expressed in terms of user-facing concepts, incremental refreshes should be treated strictly as a performance optimization, and the system should gracefully handle changes to upstream dependencies. 

We implemented Dynamic Tables by building atop Snowflake's existing catalog, scheduling system, transaction engine, query optimizer, and query processor, modifying each only where needed. The result is that Dynamic Tables implements micro-batch processing~\cite{microbatching}, where each micro-batch is an optimized, relational query plan executed on Snowflake's vectorized, push-based query processor running in the context of Snowflake's transaction engine.
In this section, we describe some extensions we made to these components to implement Dynamic Tables.

\subsection{Architecture}
\begin{figure}[h]
  \centering
  \includegraphics[width=\linewidth]{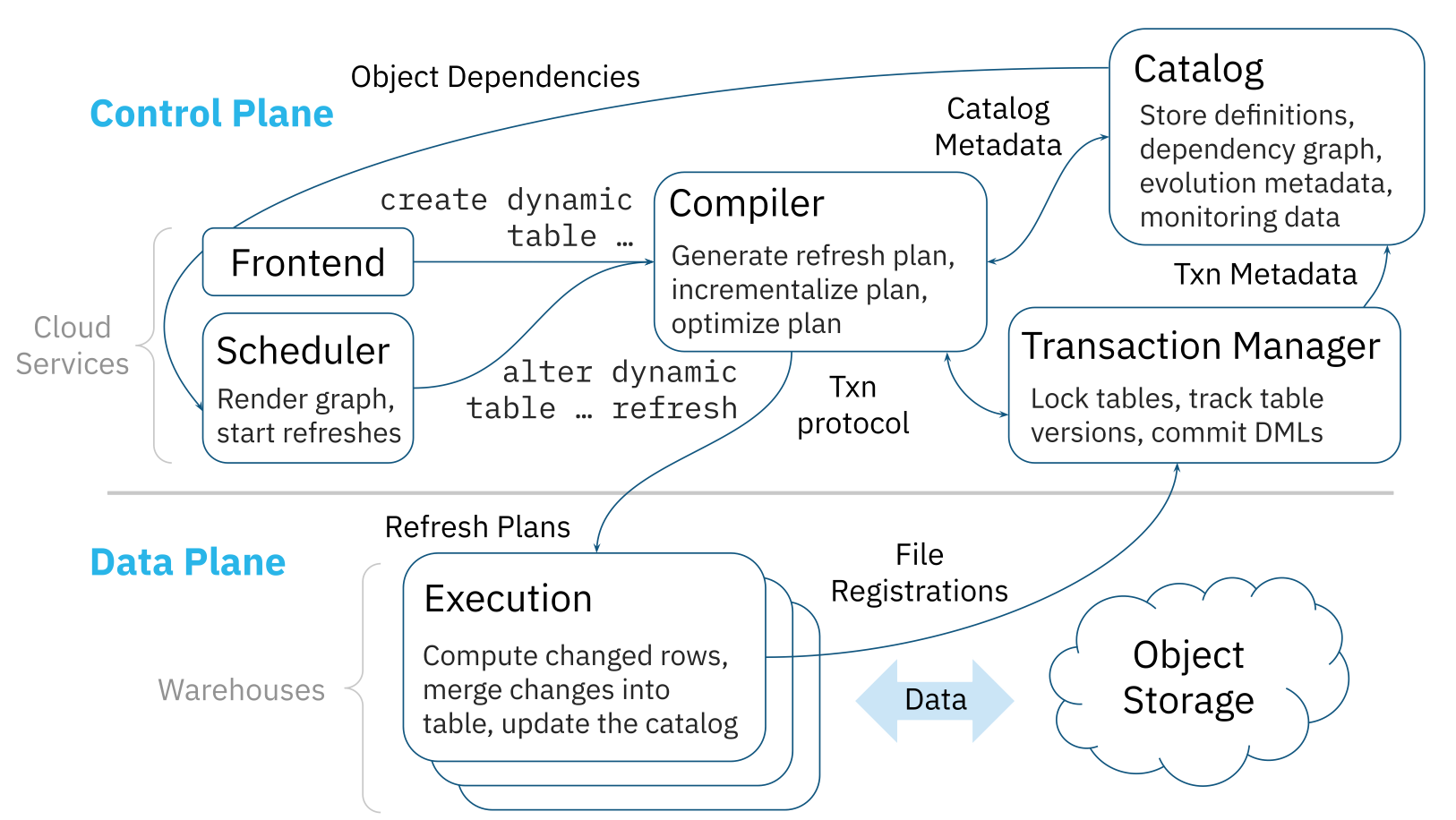}
  \caption{An architecture diagram of Dynamic Tables.}
  \label{fig:arch}
\end{figure}

The architecture of Dynamic Tables mirrors the architecture of Snowflake, as shown in Figure~\ref{fig:arch}. Queries arrive at the frontend; they proceed into the compiler, which parses, binds identifiers, and generates an optimized query plan. The catalog stores the metadata needed by the compiler. Many query plans are executed on a warehouse, which communicates with the control plane about query status and modifications to tables. Table data is stored in object storage, whence it is fetched and cached by warehouse nodes. The transaction manager handles versioning of table metadata, manages locks, tracks uncommitted changes, and atomically commits transactions. The scheduler works in the background, performing asynchronous maintenance tasks as needed.

Dynamic Tables extends all of these components. Straightforward extensions to the compiler and catalog were developed to support the many DDL commands required to manage Dynamic Tables. The bulk of the extensions were developed in support of scheduling, executing, and monitoring refresh operations. The catalog generates a timestamped, linearizable log of DDL operations to all DTs and related entities. This DDL log is consumed by a job in the scheduler that renders the dependency graph of DTs and issues refresh commands as required to meet the target lag of each. Each refresh command is sent to the compiler, which rewrites the DT's defining query into an optimized query plan, making it incremental if applicable. This rewrite process looks up dependencies in the catalog, applying repairs as needed to maintain compatibility. Appropriate table versions are resolved to ensure that snapshot isolation is preserved by the refresh. The plan is sent to a virtual warehouse, which executes a standard Snowflake query plan with minimal modifications. The execution of this plan is coordinated with the transaction manager, which locks the DT, stages changes to its contents, commits or rolls back those changes, creates a new table version indexed by the data timestamp, and unlocks the table.

\subsection{Scheduling}
\begin{figure}[h]
  \centering
  \includegraphics[width=\linewidth]{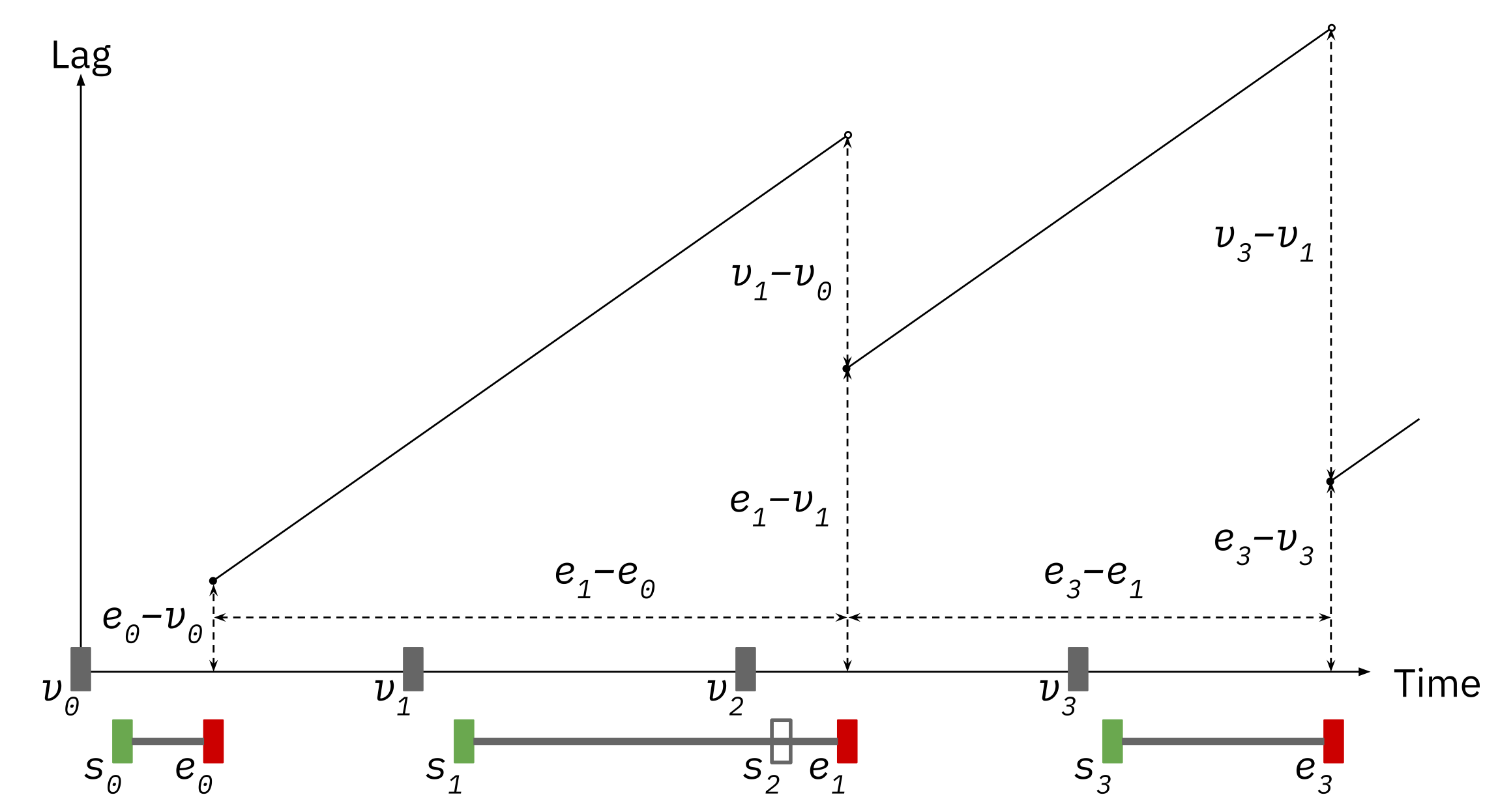}
  \caption{An example of lag over time. Boxes $v_i$ are data timestamps. $s_i$ and $e_i$ denote the start and end times of refreshes. Dashed arrows are measurements.}
  \label{fig:lagdiagram}
\end{figure}

Snowflake schedules refreshes of each DT in order to try to meet its target lag. This takes the form of a sequence of refreshes for each DT, each of which has a data timestamp, a start time, and an end time. 
Given a sequence of refreshes, the lag is a sawtooth that rises at a constant rate of 1~second per second, as shown by the example in Figure~\ref{fig:lagdiagram}. 

\label{sec:sched}

When a refresh commits, the lag instantaneously drops, forming a sawtooth with a peak and trough at the end time. The lag at a trough is the end time of that refresh minus its data timestamp. 
For example, for refresh~1, the trough lag is $e_1 - v_1$. The lag at a peak is the end time of that refresh minus the data timestamp of the \emph{preceding} refresh. For example, for refresh~1, the peak lag is $e_1 - v_0$ since until refresh~1 actually commits, the data in the table is up to date only as of $v_0$. This leads to the somewhat counterintuitive conclusion that, when scheduling a refresh to try to meet a target lag, one needs to consider the data timestamp of the preceding refresh, not just the expected duration of the refresh to be scheduled. 

Another way to see this is by splitting the peak lag into 3 elements: $p$ is the duration between the data timestamps of the adjacent refreshes, $w$ is the waiting time between the data timestamp and the start time of the refresh, and $d$ is the duration of the refresh. Staying within a target lag $t$ requires:
\[p + w + d < t\] 
$p$ can be thought of as the period between refreshes, though this period may not be constant. $w$ is complicated by the presence of upstream DTs, and we discuss it further in the next paragraph. $d$ is a consequence of the amount of work required by the refresh and the resources allocated to it. Snowflake has full control over $p$, some control over $w$, and limited control over $d$.

Scheduling refreshes across the graph of DTs introduces an additional challenge. A refresh of a DT at a given data timestamp cannot begin until all of its upstream DTs have refreshed at the same data timestamp. This means that, for all DTs $i$: 
\[w_i \ge \max(w_j + d_j) \ \forall \ j \in \text{upstream}(i)\]
Note that this formula is recursive because each $w_j$ is itself bounded below by the refresh durations of DTs further upstream.

Snowflake's primary control for keeping DTs within their target lag is the choice of $p$ for each DT. Making $p$ too small increases the cost of the DT. Making $p$ too large runs the risk of exceeding the target lag. Therefore, the choice of $p$ must balance these two, subject to the constraints above. This is a complex optimization problem that could be solved in a rigorous way, using estimates and history to decide exactly when to refresh each DT. 

We chose to implement a simple heuristic. Recall that snapshot isolation requires that, before refreshing a DT at a certain data timestamp, all upstream DTs must also refresh at that data timestamp. This constraint means that Snowflake's freedom to vary $p$ within a connected component of DTs is quite limited. All DTs in that component are frequently forced to refresh at the same data timestamp, even if they have different target lags. Our heuristic takes advantage of this fact. 
We define a set of canonical refresh periods as $48 \cdot 2^n$ seconds, for integers $n$. When deciding upon the refresh period for a DT, we choose from this set of canonical periods to try to keep each DT within its target lag. We also ensure that the choice of refresh period for each DT is greater than or equal to those upstream. Because powers of two are all multiples of each other and we choose a constant phase for each customer, the data timestamps of different DTs are guaranteed to align, even if they have different target lags.

This heuristic has been adequate for most Snowflake customers, but it does have limitations. First, it occasionally causes confusion for new users because the refresh period Snowflake chooses can be substantially smaller than the provided target lag. Some users unknowingly conflate the refresh period and the target lag, and are surprised to discover this discrepancy. Second, for graphs of DTs with long chains or highly variable workloads, this approach limits the responsiveness of the system, as we can only double or halve the refresh period. We aim to explore alternative approaches in the future which choose data timestamps dynamically by analyzing the dependency graph and previous refresh durations. We will also be implementing adaptive resource scaling, which will give Snowflake more control over both $w$ and $d$, making it easier to stay within target lag even for challenging workloads.

\subsection{Transaction and Version Management}
The scheduler initiates a refresh by issuing an internal command to Snowflake, specifying a DT and a \emph{refresh timestamp}, which is the DT's new data timestamp after the refresh commits. This command acts like a DML operation within a transaction. 
We focus on 3 key aspects related to transactions: resolving appropriate table versions, tracking progress, and managing conflicts.

Resolving the correct table version is instrumental to upholding the Snapshot Isolation guarantee. 
When Snowflake creates a new table version, its visibility is determined by the commit timestamp of the transaction that created it. This timestamp is read from a Hybrid Logical Clock (HLC)~\cite{hlc}, and is totally ordered relative to the commits of all other transactions in the account. When trying to read a table version as of some time $t$, Snowflake looks up the table version with the largest commit timestamp less than or equal to $t$. A DT refresh applies this mechanism to resolve the version of regular tables as of the refresh timestamp.
However, when a DT $d$ reads from another DT $u$, we need to find the table version of $u$ that was created by a refresh of $u$ with the same refresh timestamp as the current refresh of $d$. There can be a significant delay between a table version's commit timestamp and its corresponding refresh timestamp. So, we store a mapping from refresh timestamp to commit timestamp for each DT's table versions. When a refresh commits, we add a new entry to the mapping; to look up a version for a particular refresh timestamp, we consult the mapping.

Each time a DT refreshes, its data timestamp moves forward in time. But the data timestamp is an abstraction over a more complicated object we call a \emph{frontier}. A frontier is a map containing the table version of each source table that the DT has consumed, and an HLC timestamp of that refresh. There are several cases the require tracking each individual source. For example, if there are bugs in how versions are tracked, frontiers provide much more debugging information for investigation. The other cases relate to advanced features like cloning, replication, and sharing, and we do not discuss them here.

During compilation, we look up the DT's current frontier and generate a new frontier from the refresh timestamp. These two frontiers comprise the interval over which the refresh will advance. Based on this interval, we decide which refresh plan to use (\S\ref{sec:rewrite}). When committing a refresh, we overwrite the old frontier with the new one to mark the progress of the DT.

Conflicts are managed using locks. Each Dynamic Table is locked when a refresh operation begins, and unlocked after it commits. There are opportunities to run some refreshes concurrently. For example, two refreshes which only insert data do not conflict. We expect to take advantage of these opportunities as we push the minimum target lag to lower values.

All of the above functionality reuses code from Snowflake's Tables and Streams and supports all of the same monitoring and debugging tools, which has substantially accelerated our development.

\subsection{Query Rewrites}
\label{sec:rewrite}
Recall that the scheduler initiates a refresh by issuing an internal command to Snowflake specifying the DT to refresh and its data timestamp. The compiler pipeline rewrites a syntax tree containing only these two parameters into a refresh plan in several steps. To begin, the defining query of the DT is expanded and inserted into the syntax tree beneath the node representing the refresh command. Identifiers in this tree are bound and nested views are expanded. This expanded syntax tree is then translated into a relational representation, on which further rewrites are performed.

We first check if the definition of the DT has changed.
When a DT is created, we track all of its dependencies and store them as metadata for the DT. Dependencies can be entities referred to directly as part of the definition, or indirectly, such as objects inside of views or data access policies. For each table-like dependency, we also track the specific columns that are used by the DT. 
During a refresh, the DT may have different columns (e.\ g., for a top-level \code{SELECT *}) or altogether different semantics (e.\ g., changing a filter or reading from a different table) due to DDLs on objects upstream.
Query evolution determines how to compensate for the changes, whether via DDL actions or overriding the refresh action.
Our approach is currently  conservative, choosing to reinitialize in some cases where it is not necessary. In the future, we would like to make these checks more selective, add support for automatic schema evolution, and implement partial reinitializations for cases that do not require recomputing the whole DT again.

Next, we determine the refresh action to take. The \code{NO\_DATA} action is taken when the sources of the DT have not changed since the last refresh's data timestamp. We determine this by looking at the metadata and version history of the underlying tables. For this action, we merely commit a transaction marking the progress of the DT to the next data timestamp. This uses neglible resources and zero Virtual Warehouse compute.

If the sources of the DT have changed, and the DT has \code{FULL} refresh mode, the \code{FULL} action is chosen. In this case, we add an \code{INSERT OVERWRITE} plan node at the top of the query and set the table versions used in table scans to those corresponding to the next data timestamp.

When the DT has an \code{INCREMENTAL} refresh mode, it applies the \code{INCREMENTAL} action unless query evolution forces it to \code{REINITIALIZE}. An incremental refresh computes the changes in the query between the previous and next data timestamps and applies those changes to the stored data. A reinitialization is similar to a full refresh, except that it also computes metadata that is required for incremental refreshes.

Once the choice of action is made, the plan is rewritten accordingly. For the \code{INCREMENTAL} action, we add the query differentiation operator at the top of the query plan to instruct the next phase to rewrite it into an incremental plan. Then, we add a merge operator on top of that, which applies the \code{DELETE} and \code{INSERT} actions to the DT itself. For the \code{REINITIALIZE} action, we add the initial-condition operator to instruct the next phase to define row IDs throughout the plan. Then, we add an insert-overwrite operator on top, which includes those row IDs in addition to the user-defined columns.

\subsection{Query Differentiation}
\label{sec:incr}
In~\cite{akidau2023difference}, we described the extensible query differentiation framework we developed in support of Snowflake Streams on Views. Dynamic Tables reuses this framework and extends it in support of new operators and optimizations. We provide a short summary here and refer the reader to the original paper for more details.

Incremental DTs define a unique ID for every row in the query result, and store those IDs alongside the data. To perform an incremental refresh, Snowflake differentiates the DT's defining query $Q$ to produce a query $\DI Q$ that outputs the changes in that query over a data timestamp interval $I$. These changes are a set of rows with the same columns as $Q$, plus 2 additional metadata columns. The \code{\$ACTION} column indicates whether a row represents an insertion or a deletion in the DT. Updates are represented as both actions for the same row. The \code{\$ROW\_ID} column provides the identifier of the row to be modified. The differentiation framework guarantees that a set of changes never contains more than 1 row for each unique \code{\$ROW\_ID}, \code{\$ACTION} pair, which ensures that the merge operation is well-defined.

The framework is implemented in terms of syntactic rewrite rules, which match the derivative operator and the plan beneath it, and produce an equivalent expression in terms of derivatives of its internal terms.
As long as the plan only contains differentiable operators, this process eliminates all derivatives, resulting in a plan that contains only executable operators like scan, project, filter, join, etc.
After being rewritten, the plan is optimized and executed like any other.

\subsubsection{More Derivatives}
For DTs, we implemented differentiation for several additional operators. Here, we focus on two in particular: outer joins and window functions. In our original implementation of outer join derivatives, we took advantage of the fact that an outer join is logically equivalent to an inner join and one or two anti-joins:
\[
\DI(Q \leftouterjoin R) \Longrightarrow 
        \DI(Q \bowtie R)  +  \DI(\pi_{R=\code{NULL}}(Q \vartriangleright R)) 
\]

By rewriting outer joins in this way, we are able to differentiate outer joins by reusing our inner join and anti-join derivatives. This approach worked, but it had undesirable performance characteristics due to the repetition of the $Q$ and $R$ terms. 
Snowflake supports term reuse via a \code{SPLIT} operator, which is used to implement common table expressions. However, some optimizations are incompatible with reused terms. This left us with the difficult choice between giving up optimizations or having duplicate subplans in the query, where the duplication grows exponentially with the number of outer joins in the plan.

To address this problem, we implemented a direct differentiation operator for outer joins, factoring out common terms in a way that sidestepped the aforementioned problems. The resulting algebra is messy and not particularly illuminating, so we omit it for brevity. The takeaway is that algebraic choices that seem mathematically trivial can interact with the optimizer and other parts of the query processing engine, imposing a substantial impact on performance when differentiating queries.

Snowflake's support for window functions includes partitioned and unpartitioned input, ordered and unordered aggregation, row-based and range-based frames, and cumulative and sliding frames. Different combinations of these features can be implemented in many ways, with complex performance trade-offs~\cite{leis}. We believe that these trade-offs become even more complex when trying to implement differentiation on them. Rather than tackle all of this complexity at once, we decided to take a simplistic approach: define a single derivative that is as general as possible but leaves some performance on the table. This derivative works by applying the window function to all partitions that have changed. This is the rule, where $\xi_k$ represents a window function partitioned by keys $k$:
\[
    \DI(\xi_k(Q)) \Longrightarrow \pi_- (\xi_k(Q|_{I_0} \ltimes_k \DI Q)) + \pi_+ (\xi_k(Q|_{I_1} \ltimes_k \DI Q))
\]

This derivative does not reuse any of the work from previous refreshes within a single-window partition. However, it works for all window functions with \code{PARTITION BY} clauses (as long as ties in \code{ORDER BY} are broken repeatably). We expect to implement further optimizations for specific, common window functions, an area that deserves more research.

\subsubsection{More Optimizations}
Beyond these two additional derivatives, we implemented a number of optimizations to improve the performance of incremental refreshes. Unlike the row IDs that we generate for streams, which are \code{SHA-1} hashes, the row IDs we use inside of Dynamic Tables contain plaintext prefixes to improve the performance of joins using row IDs as a key, increasing the selectivity of Snowflake's runtime pruning. We are exploring different designs for our representation of row IDs to reduce the number of hash operations we perform and improve compressibility.

Although our framework is able to gracefully handle arbitrary changes in source data, substantial performance boosts are available if the framework is allowed to take advantage of constraints on those changes. Insert-only workloads are extremely common, and specializing a plan to work only with inserts can provide substantial speedups. In many cases, the structure of a query guarantees that redundant actions will not be introduced by differentiation, which permits us to skip the final change-consolidation step (see~\cite{akidau2023difference}) that ensures at most one \code{\$ROW\_ID}, \code{\$ACTION} in each result. Due to the copy-on-write nature of Snowflake's tables, naively reading from added and removed partitions, as described in the Streams paper, often causes read amplification. Depending on the amount of read amplification and the structure of the query downstream, it can be beneficial to eliminate these copied rows early on, rather than waiting until the end of the plan.

Snowflake supports automatically clustering and defragmenting data in the background to ensure stable query performance even as data distributions shift. These maintenance operations do not change the logical contents of a table, but they do add and remove files. A naive differentiation would read all of these files, only to produce an empty set of changes. Dynamic Tables tries to skip these data-equivalent operations. However, when exploring solutions to finding the optimal set of such operations to skip, we ended up concluding that it is an NP-Hard graph optimization problem. For the moment, we carved out a tractable portion of the problem, but believe this problem deserves further scrutiny.

\subsubsection{Future Work} 
Optimizing incremental queries is fertile ground for future development of Dynamic Tables. There are several large problems that we would like to investigate. First, none of our derivatives so far reuse the state from preceding data timestamps already stored in the DT. They all work by computing changes purely in terms of the sources. We expect major performance opportunities from incorporating a ``previous state'' into our differentiation rules. Second, a DT is not currently able to maintain intermediate state to accelerate incremental refreshes. We rely on customers to factor their queries into simpler fragments, but this can be toilsome. We intend to automatically split queries into fragments, with hidden, internal DTs containing the intermediate state. Third, incorporating cost-based optimization into query differentiation becomes increasingly important as the diversity of incrementalizations increases. We expect to tackle these problems in the near future, but we believe the field is a rich area for in-depth research and ripe for contributions from academia.

\section{Learnings in Production} \label{sec:production}
In this section, we relay our learnings from operating an incremental view maintenance-based streaming system at scale, covering our approach to safety and liveness, how our design decisions fared in the market, and our assessment of IVM's suitability as a foundation for stream processing.

\subsection{Safety}
\label{sec:safe}
Snowflake's engineering culture takes correctness very seriously. We design each feature carefully, build safety mechanisms into the implementation wherever possible, and do extensive testing and validation. All of this effort has paid off repeatedly in Dynamic Tables as we caught and fixed subtle, unexpected bugs. This experience reinforces the lesson that it is difficult to take safety too seriously in a DBMS. This safety is best achieved by a strategy of defense-in-depth, with each level requiring substantial investment in tooling. In the following section, we will discuss Snowflake's approach to testing, and how we applied and augmented it for Dynamic Tables.

The first level of testing is unit tests, covering things like parsing and complex stateless functions. Unfortunately, because so much of the functionality in Dynamic Tables spans across components, the amount of coverage we can get from unit tests is limited.

The second level we call \emph{integration tests} (though they differ from the typical definition). These tests are still in-memory, but they exercise logic end to end by running the whole Snowflake control plane. This layer permits testing invariants in the compilation pipeline. Thanks to liberal assertions, these tests provide significant coverage by simply compiling a wide variety of queries and ensuring the absence of errors. For example, we ensure that our transaction engine resolves table versions and locks tables as expected, and compatibility between our rewrites and other phases of the compiler.

The third level we call \emph{regression tests}. These are Python scripts that execute SQL commands against a full Snowflake stack and verify that the returned results are correct. We use this layer to test the end-to-end functionality of Dynamic Tables commands, the scheduler, and refreshes. We run them as part of our continuous integration, which mirrors our development environment, as well as during release validation, which mirrors our production environment.

The fourth level of testing is a miscellaneous bucket that we call \emph{workload tests}. These tests run outside of our core CI pipeline, and include performance tests, traditional integration tests, and property-based randomized testing. For Dynamic Tables, this last kind of test has been especially valuable. Because of delayed-view semantics with snapshot isolation, we have an extremely strong assertion we can make for most DTs: if you run the defining query as of the data timestamp, you should get the same result as in the DT. Checking this assertion within a framework that generates random SQL queries allows us to test the correctness of hundreds of thousands of different DTs in a matter of hours. We run this workload test daily and alert the team when it discovers a discrepancy.

The fifth level is \emph{Snowtrail}~\cite{snowtrail}, which allows us to re-run a customer query on two different system configurations and compare obfuscated results. We test the correctness and performance of our changes on a realistic distribution of queries, complementing the synthetic distribution generated by randomized property testing.

The sixth level of testing is a set of \emph{validations} that we run in production to catch inconsistencies. These are similar to assertions, but more sophisticated and intentional. For Dynamic Tables, we have 3 core validations. First, when a DT resolves the table version for a DT upstream, it looks for an exact version corresponding to the data timestamp of the refresh. If this version cannot be found, we fail the refresh, preventing violations of snapshot isolation and alerting us to a bug in the scheduler. Second and third, incremental DT refreshes maintain two invariants: there should never be more than 1 row with the same \code{\$ROW\_ID}, \code{\$ACTION} pair, and we should never try to delete a row that does not exist. We check these invariants for every incremental DT refresh, and fail the refresh if they are violated. This has on numerous occasions shielded customers from data corruption and alerted us to some of the most subtle, rare bugs we have ever encountered, including some that had been latent in Snowflake for years.

We continue to invest in tooling and tests to further increase our coverage and the confidence our customers can have in Snowflake.

\subsection{Liveness}
We ensure liveness with industry best-practices. We define internal SLOs that make a distinction between Snowflake's responsibilities and customer responsibilities. For example, we cannot simply assert that all DTs stay within their target lag some fraction of the time: customers control the query, the data, and the resources available. Instead, we instrumented the system so that we can always determine which state a DT is expected to be in. For example, every DT refresh emits heartbeats as long as it is running, and we have a background service that confirms that every DT that is in the \code{EXECUTING} state sent a heartbeat recently. We have alerts that notify us when the system violates one of these SLOs and an on-call rotation to respond accordingly.

\subsection{Adoption and Usage}
Dynamic Tables has been Generally Available (GA) since April 2024, and is in use by thousands of Snowflake customers, with over 1 million active tables as of 2025-04-01. Currently, almost 70\% of active DTs have an incremental refresh mode, a fraction that grows as we add support for more operators. Even so, we have found that many customers are satisfied with full refreshes.

More than 90\% of refreshes have no data, reflecting that customers often set the target lag lower than their data refresh rate. We encourage this pattern, as these refreshes are inexpensive and we intend to further reduce their cost.

\begin{figure}[h]
  \centering
  \includegraphics[width=\linewidth]{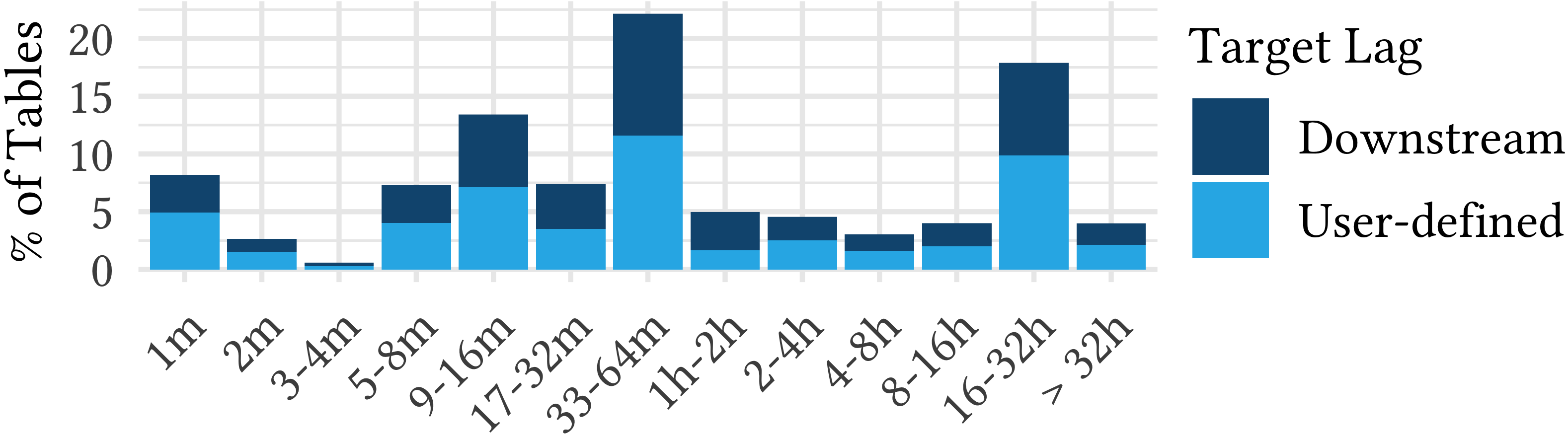}
  \caption{Distribution of the target lags of active DTs.}
  \label{fig:lag}
\end{figure}

Figure \ref{fig:lag} shows the distribution of the target lag across all DTs. More than 25\% of DTs have a target lag of at least 16 hours, firmly in the batch domain. In the streaming domain, nearly 20\% of DTs have a target lag less than 5 minutes. 
The 55\% of DTs between these validates our hypothesis that the middle ground between classic batch and streaming is underserved.

A majority (67\%) of incremental refreshes (non-initial, non-empty, scheduled) has a number of output changed rows (inserts + deletes) of less than 1\% of the total size of the respective DT, which underscores the importance of efficient incremental refreshes. 21\% of refreshes change more than 10\% of their DT, highlighting the need to be able to dynamically choose full refreshes when a large fraction of the data has changed. 

\begin{figure}[h]
  \centering
  \includegraphics[width=1\linewidth]{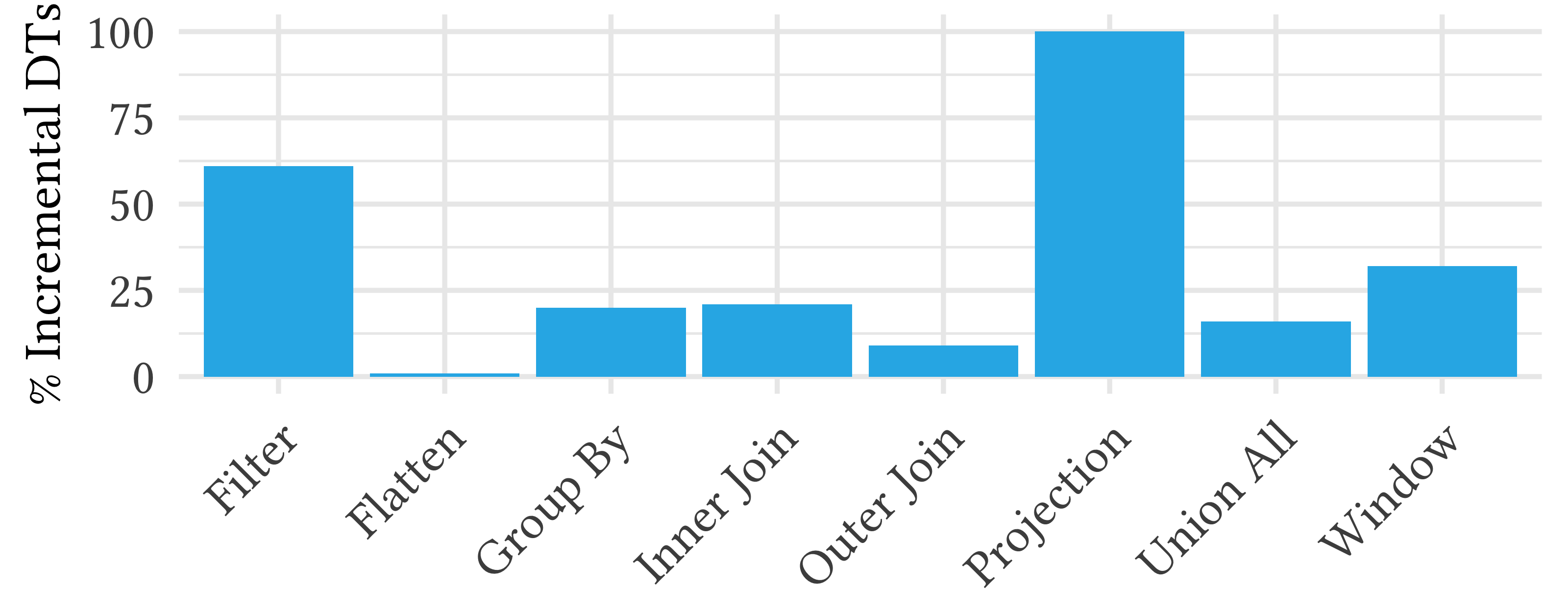}
  \caption{The frequency of each operator in the definition of incremental DTs. }
  \label{fig:features}
\end{figure}

Figure \ref{fig:features} shows the frequency of operators used in incremental DT definitions, demonstrating that joins, aggregates, and window functions are common.

Dynamic Tables supports the breadth of Snowflake features. More than 20\% of active DTs were cloned from another, and ~20\% are in a shared database. Several thousand Iceberg DTs are currently active, less than a month after their GA.

The above usage is compelling evidence that a product like Dynamic Tables can make stream processing accessible to a broad user base across a wide range of latencies.

\subsection{DVS/IVM for Streaming}
Dynamic Tables embrace delayed view semantics with incremental view maintenance (DVS/IVM) as the conceptual foundation for a stream processing product. Based on our experience, we contend that DVS/IVM is a solid foundation and a significant contributor to the success of Dynamic Tables. However, DVS/IVM presents inherent challenges that remain unresolved.

IVM’s strengths lie in its simplicity and accessibility. Users familiar with SQL can easily create DT pipelines, and the rise of frameworks like \texttt{dbt}~\cite{dbt}, which promote view-like semantics, has cultivated a ready user base. Additionally, as discussed in \S\ref{sec:safe}, DVS/IVM’s predictable nature simplifies testing, offering clear benefits over streaming systems without similar foundations.

Despite these advantages, DVS/IVM has notable limitations. In some cases, even the best incremental performance is insufficient. For example, updating a dimension table in a star schema that joins with many facts can be as costly as rewriting the entire table. In such scenarios, consistent performance often outweighs semantic purity, yet expressing this intent through SQL typically introduces undue complexity. Furthermore, DVS/IVM struggles with workflows that extend beyond view semantics. Data expiration is one such example: it is common to refine raw data into another table and discard the raw data after a period, but ensuring that expired data does not affect results is challenging. Similarly, use cases requiring irrevocable actions challenge DVS/IVM’s assumption of mutability. Temporary delays in metric delivery, for instance, may trigger spurious alerts in a telemetry system. Real-world scenarios often demand waiting until results are final \cite{watermarks}.

We remain confident in our commitment to DVS/IVM as a foundation for streaming. However, addressing these limitations will require both theoretical and practical advancements. DVS/IVM provides a robust basis for tackling these challenges, and we aim to share our solutions in the future. We invite the DBMS and Streaming communities to join the discussion to advance this critical area.

\section{Conclusion} \label{sec:conclusion}

In this paper, we presented delayed view semantics and Dynamic Tables. With delayed view semantics, we extend the existing body of transaction isolation research to formalize the notion of tables whose contents present a consistent view on data at some time in the past. We do this by introducing the idea of a derivation, which embodies the fact that pure computation can be moved between transactions arbitrarily without affecting application invariants. This enables us to reason precisely about the transactional phenomena induced by asynchronous, pure computation, and thereby solve one of the remaining open problems in bridging the domains of stream processing and databases.

With Dynamic Tables, we package delayed view semantics in a simple, declarative transformation primitive built into the Snowflake analytical RDBMS. Users specify their transformations as a SQL query and an associated lag, and Dynamic Tables takes care of everything else: incrementalization, scheduling, execution, and integration with Snowflake's broad set of enterprise-ready features. We discuss the implementation, detailing the challenges faced and benefits gained by building this feature deep into an existing world class analytical database. This work yielded a product that has a level of accessibility unprecedented in a stream processor, and has seen a corresponding uptake across our customer base.

We hope that this work paves the way for a broader convergence of stream processors and databases, which we believe can unlock major opportunities for developer experience and performance.
DVS/IVM provide a rigorous foundation for this convergence, but many theoretical and practical problems remain to be solved, such as how to best implement an efficient, scalable IVM engine, and how to transcend the semantic limitations of DVS.
We are excited about the possibilities awaiting the field as we rise to meet these challenges.

\input{output.bbl}

\end{document}

%% file: output.bbl

%% file: paper.bbl
\begin{thebibliography}{34}


\ifx \showCODEN    \undefined \def \showCODEN     #1{\unskip}     \fi
\ifx \showDOI      \undefined \def \showDOI       #1{#1}\fi
\ifx \showISBNx    \undefined \def \showISBNx     #1{\unskip}     \fi
\ifx \showISBNxiii \undefined \def \showISBNxiii  #1{\unskip}     \fi
\ifx \showISSN     \undefined \def \showISSN      #1{\unskip}     \fi
\ifx \showLCCN     \undefined \def \showLCCN      #1{\unskip}     \fi
\ifx \shownote     \undefined \def \shownote      #1{#1}          \fi
\ifx \showarticletitle \undefined \def \showarticletitle #1{#1}   \fi
\ifx \showURL      \undefined \def \showURL       {\relax}        \fi
\providecommand\bibfield[2]{#2}
\providecommand\bibinfo[2]{#2}
\providecommand\natexlab[1]{#1}
\providecommand\showeprint[2][]{arXiv:#2}

\bibitem[Adya(1999)]%
        {adya1999weak}
\bibfield{author}{\bibinfo{person}{Atul Adya}.} \bibinfo{year}{1999}\natexlab{}.
\newblock \showarticletitle{Weak consistency: a generalized theory and optimistic implementations for distributed transactions}.
\newblock  (\bibinfo{year}{1999}).
\newblock


\bibitem[Adya et~al\mbox{.}(2000)]%
        {adya}
\bibfield{author}{\bibinfo{person}{A. Adya}, \bibinfo{person}{B. Liskov}, {and} \bibinfo{person}{P. O'Neil}.} \bibinfo{year}{2000}\natexlab{}.
\newblock \showarticletitle{Generalized isolation level definitions}. In \bibinfo{booktitle}{\emph{Proceedings of 16th International Conference on Data Engineering (Cat. No.00CB37073)}}. \bibinfo{pages}{67--78}.
\newblock
\urldef\tempurl%
\url{https://doi.org/10.1109/ICDE.2000.839388}
\showDOI{\tempurl}


\bibitem[Ahmad et~al\mbox{.}(2012)]%
        {dbtoaster}
\bibfield{author}{\bibinfo{person}{Yanif Ahmad}, \bibinfo{person}{Oliver Kennedy}, \bibinfo{person}{Christoph Koch}, {and} \bibinfo{person}{Milos Nikolic}.} \bibinfo{year}{2012}\natexlab{}.
\newblock \showarticletitle{DBToaster: higher-order delta processing for dynamic, frequently fresh views}.
\newblock \bibinfo{journal}{\emph{Proc. VLDB Endow.}} \bibinfo{volume}{5}, \bibinfo{number}{10} (\bibinfo{date}{June} \bibinfo{year}{2012}), \bibinfo{pages}{968–979}.
\newblock
\showISSN{2150-8097}
\urldef\tempurl%
\url{https://doi.org/10.14778/2336664.2336670}
\showDOI{\tempurl}


\bibitem[Akidau et~al\mbox{.}(2013)]%
        {akidau2013millwheel}
\bibfield{author}{\bibinfo{person}{Tyler Akidau}, \bibinfo{person}{Alex Balikov}, \bibinfo{person}{Kaya Bekiro{\u{g}}lu}, \bibinfo{person}{Slava Chernyak}, \bibinfo{person}{Josh Haberman}, \bibinfo{person}{Reuven Lax}, \bibinfo{person}{Sam McVeety}, \bibinfo{person}{Daniel Mills}, \bibinfo{person}{Paul Nordstrom}, {and} \bibinfo{person}{Sam Whittle}.} \bibinfo{year}{2013}\natexlab{}.
\newblock \showarticletitle{Millwheel: Fault-tolerant stream processing at internet scale}.
\newblock \bibinfo{journal}{\emph{Proceedings of the VLDB Endowment}} \bibinfo{volume}{6}, \bibinfo{number}{11} (\bibinfo{year}{2013}), \bibinfo{pages}{1033--1044}.
\newblock


\bibitem[Akidau et~al\mbox{.}(2023)]%
        {akidau2023difference}
\bibfield{author}{\bibinfo{person}{Tyler Akidau}, \bibinfo{person}{Paul Barbier}, \bibinfo{person}{Istvan Cseri}, \bibinfo{person}{Fabian Hueske}, \bibinfo{person}{Tyler Jones}, \bibinfo{person}{Sasha Lionheart}, \bibinfo{person}{Daniel Mills}, \bibinfo{person}{Dzmitry Pauliukevich}, \bibinfo{person}{Lukas Probst}, \bibinfo{person}{Niklas Semmler}, \bibinfo{person}{Dan Sotolongo}, {and} \bibinfo{person}{Boyuan Zhang}.} \bibinfo{year}{2023}\natexlab{}.
\newblock \showarticletitle{What's the Difference? Incremental Processing with Change Queries in Snowflake}.
\newblock \bibinfo{journal}{\emph{Proc. ACM Manag. Data}} \bibinfo{volume}{1}, \bibinfo{number}{2}, Article \bibinfo{articleno}{196} (\bibinfo{date}{June} \bibinfo{year}{2023}), \bibinfo{numpages}{27}~pages.
\newblock
\urldef\tempurl%
\url{https://doi.org/10.1145/3589776}
\showDOI{\tempurl}


\bibitem[Akidau et~al\mbox{.}(2021)]%
        {watermarks}
\bibfield{author}{\bibinfo{person}{Tyler Akidau}, \bibinfo{person}{Edmon Begoli}, \bibinfo{person}{Slava Chernyak}, \bibinfo{person}{Fabian Hueske}, \bibinfo{person}{Kathryn Knight}, \bibinfo{person}{Kenneth Knowles}, \bibinfo{person}{Daniel Mills}, {and} \bibinfo{person}{Dan Sotolongo}.} \bibinfo{year}{2021}\natexlab{}.
\newblock \showarticletitle{Watermarks in stream processing systems: semantics and comparative analysis of Apache Flink and Google cloud dataflow}.
\newblock \bibinfo{journal}{\emph{Proc. VLDB Endow.}} \bibinfo{volume}{14}, \bibinfo{number}{12} (\bibinfo{date}{July} \bibinfo{year}{2021}), \bibinfo{pages}{3135–3147}.
\newblock
\showISSN{2150-8097}
\urldef\tempurl%
\url{https://doi.org/10.14778/3476311.3476389}
\showDOI{\tempurl}


\bibitem[Akidau et~al\mbox{.}(2015)]%
        {akidau2015dataflow}
\bibfield{author}{\bibinfo{person}{Tyler Akidau}, \bibinfo{person}{Robert Bradshaw}, \bibinfo{person}{Craig Chambers}, \bibinfo{person}{Slava Chernyak}, \bibinfo{person}{Rafael~J Fern{\'a}ndez-Moctezuma}, \bibinfo{person}{Reuven Lax}, \bibinfo{person}{Sam McVeety}, \bibinfo{person}{Daniel Mills}, \bibinfo{person}{Frances Perry}, \bibinfo{person}{Eric Schmidt}, {et~al\mbox{.}}} \bibinfo{year}{2015}\natexlab{}.
\newblock \showarticletitle{The dataflow model: a practical approach to balancing correctness, latency, and cost in massive-scale, unbounded, out-of-order data processing}.
\newblock \bibinfo{journal}{\emph{Proceedings of the VLDB Endowment}} \bibinfo{volume}{8}, \bibinfo{number}{12} (\bibinfo{year}{2015}), \bibinfo{pages}{1792--1803}.
\newblock


\bibitem[Amazon Web~Services({[n.\,d.]})]%
        {site-redshift-mv}
\bibfield{author}{\bibinfo{person}{Inc. Amazon Web~Services}.} \bibinfo{year}{[n.\,d.]}\natexlab{}.
\newblock \bibinfo{title}{Redshift Materialized Views}.
\newblock \bibinfo{howpublished}{\url{https://docs.aws.amazon.com/redshift/latest/dg/materialized-view-overview.html}}.
\newblock
\newblock
\shownote{Accessed: 2024-12-02}.


\bibitem[Armbrust et~al\mbox{.}(2018)]%
        {armbrust2018structured}
\bibfield{author}{\bibinfo{person}{Michael Armbrust}, \bibinfo{person}{Tathagata Das}, \bibinfo{person}{Joseph Torres}, \bibinfo{person}{Burak Yavuz}, \bibinfo{person}{Shixiong Zhu}, \bibinfo{person}{Reynold Xin}, \bibinfo{person}{Ali Ghodsi}, \bibinfo{person}{Ion Stoica}, {and} \bibinfo{person}{Matei Zaharia}.} \bibinfo{year}{2018}\natexlab{}.
\newblock \showarticletitle{Structured streaming: A declarative api for real-time applications in apache spark}. In \bibinfo{booktitle}{\emph{Proceedings of the 2018 International Conference on Management of Data}}. \bibinfo{pages}{601--613}.
\newblock


\bibitem[Bello et~al\mbox{.}(1998)]%
        {bello98mvoracle}
\bibfield{author}{\bibinfo{person}{Randall Bello}, \bibinfo{person}{Karl Dias}, \bibinfo{person}{Alan Downing}, \bibinfo{person}{James Jr}, \bibinfo{person}{James Finnerty}, \bibinfo{person}{William Norcott}, \bibinfo{person}{Harry Sun}, \bibinfo{person}{Andrew Witkowski}, {and} \bibinfo{person}{Mohamed Ziauddin}.} \bibinfo{year}{1998}\natexlab{}.
\newblock \showarticletitle{Materialized Views in Oracle.} \bibinfo{pages}{659--664}.
\newblock


\bibitem[Blakeley et~al\mbox{.}(1986)]%
        {blakely86efficient}
\bibfield{author}{\bibinfo{person}{Jose~A. Blakeley}, \bibinfo{person}{Per-Ake Larson}, {and} \bibinfo{person}{Frank~Wm Tompa}.} \bibinfo{year}{1986}\natexlab{}.
\newblock \showarticletitle{Efficiently updating materialized views}. In \bibinfo{booktitle}{\emph{Proceedings of the 1986 ACM SIGMOD International Conference on Management of Data}} (Washington, D.C., USA) \emph{(\bibinfo{series}{SIGMOD '86})}. \bibinfo{publisher}{Association for Computing Machinery}, \bibinfo{address}{New York, NY, USA}, \bibinfo{pages}{61–71}.
\newblock
\showISBNx{0897911911}
\urldef\tempurl%
\url{https://doi.org/10.1145/16894.16861}
\showDOI{\tempurl}


\bibitem[Budiu et~al\mbox{.}(2023)]%
        {budiu2022dbsp}
\bibfield{author}{\bibinfo{person}{Mihai Budiu}, \bibinfo{person}{Tej Chajed}, \bibinfo{person}{Frank McSherry}, \bibinfo{person}{Leonid Ryzhyk}, {and} \bibinfo{person}{Val Tannen}.} \bibinfo{year}{2023}\natexlab{}.
\newblock \showarticletitle{DBSP: Automatic Incremental View Maintenance for Rich Query Languages}.
\newblock \bibinfo{journal}{\emph{Proc. VLDB Endow.}} \bibinfo{volume}{16}, \bibinfo{number}{7} (\bibinfo{date}{March} \bibinfo{year}{2023}), \bibinfo{pages}{1601–1614}.
\newblock
\showISSN{2150-8097}
\urldef\tempurl%
\url{https://doi.org/10.14778/3587136.3587137}
\showDOI{\tempurl}


\bibitem[Carbone et~al\mbox{.}(2015)]%
        {carbone2015apache}
\bibfield{author}{\bibinfo{person}{Paris Carbone}, \bibinfo{person}{Asterios Katsifodimos}, \bibinfo{person}{Stephan Ewen}, \bibinfo{person}{Volker Markl}, \bibinfo{person}{Seif Haridi}, {and} \bibinfo{person}{Kostas Tzoumas}.} \bibinfo{year}{2015}\natexlab{}.
\newblock \showarticletitle{Apache flink: Stream and batch processing in a single engine}.
\newblock \bibinfo{journal}{\emph{The Bulletin of the Technical Committee on Data Engineering}} \bibinfo{volume}{38}, \bibinfo{number}{4} (\bibinfo{year}{2015}).
\newblock


\bibitem[Dageville et~al\mbox{.}(2016)]%
        {snowflake}
\bibfield{author}{\bibinfo{person}{Benoit Dageville}, \bibinfo{person}{Thierry Cruanes}, \bibinfo{person}{Marcin Zukowski}, \bibinfo{person}{Vadim Antonov}, \bibinfo{person}{Artin Avanes}, \bibinfo{person}{Jon Bock}, \bibinfo{person}{Jonathan Claybaugh}, \bibinfo{person}{Daniel Engovatov}, \bibinfo{person}{Martin Hentschel}, \bibinfo{person}{Jiansheng Huang}, \bibinfo{person}{Allison~W. Lee}, \bibinfo{person}{Ashish Motivala}, \bibinfo{person}{Abdul~Q. Munir}, \bibinfo{person}{Steven Pelley}, \bibinfo{person}{Peter Povinec}, \bibinfo{person}{Greg Rahn}, \bibinfo{person}{Spyridon Triantafyllis}, {and} \bibinfo{person}{Philipp Unterbrunner}.} \bibinfo{year}{2016}\natexlab{}.
\newblock \showarticletitle{The Snowflake Elastic Data Warehouse}. In \bibinfo{booktitle}{\emph{Proceedings of the 2016 International Conference on Management of Data}} (San Francisco, California, USA) \emph{(\bibinfo{series}{SIGMOD '16})}. \bibinfo{publisher}{Association for Computing Machinery}, \bibinfo{address}{New York, NY, USA}, \bibinfo{pages}{215–226}.
\newblock
\showISBNx{9781450335317}
\urldef\tempurl%
\url{https://doi.org/10.1145/2882903.2903741}
\showDOI{\tempurl}


\bibitem[Databricks({[n.\,d.]})]%
        {site-dlt}
\bibfield{author}{\bibinfo{person}{Inc. Databricks}.} \bibinfo{year}{[n.\,d.]}\natexlab{}.
\newblock \bibinfo{title}{Delta Live Tables}.
\newblock \bibinfo{howpublished}{\url{https://www.databricks.com/product/delta-live-tables}}.
\newblock
\newblock
\shownote{Accessed: 2024-12-02}.


\bibitem[dbt Labs({[n.\,d.]})]%
        {dbt}
\bibfield{author}{\bibinfo{person}{Inc. dbt Labs}.} \bibinfo{year}{[n.\,d.]}\natexlab{}.
\newblock \bibinfo{title}{dbt}.
\newblock \bibinfo{howpublished}{\url{https://getdbt.com}}.
\newblock
\newblock
\shownote{Accessed: 2024-12-02}.


\bibitem[Gjengset et~al\mbox{.}(2018)]%
        {gjengset2018noria}
\bibfield{author}{\bibinfo{person}{Jon Gjengset}, \bibinfo{person}{Malte Schwarzkopf}, \bibinfo{person}{Jonathan Behrens}, \bibinfo{person}{Lara~Timb{\'o} Ara{\'u}jo}, \bibinfo{person}{Martin Ek}, \bibinfo{person}{Eddie Kohler}, \bibinfo{person}{M~Frans Kaashoek}, {and} \bibinfo{person}{Robert Morris}.} \bibinfo{year}{2018}\natexlab{}.
\newblock \showarticletitle{Noria: dynamic, partially-stateful data-flow for high-performance web applications}. In \bibinfo{booktitle}{\emph{13th USENIX Symposium on Operating Systems Design and Implementation (OSDI 18)}}. \bibinfo{pages}{213--231}.
\newblock


\bibitem[Google({[n.\,d.]})]%
        {site-google-mv}
\bibfield{author}{\bibinfo{person}{Inc. Google}.} \bibinfo{year}{[n.\,d.]}\natexlab{}.
\newblock \bibinfo{title}{BigQuery Materialized Views}.
\newblock \bibinfo{howpublished}{\url{https://cloud.google.com/bigquery/docs/materialized-views-intro}}.
\newblock
\newblock
\shownote{Accessed: 2024-12-02}.


\bibitem[Griffin and Libkin(1995)]%
        {griffin95dups}
\bibfield{author}{\bibinfo{person}{Timothy Griffin} {and} \bibinfo{person}{Leonid Libkin}.} \bibinfo{year}{1995}\natexlab{}.
\newblock \showarticletitle{Incremental maintenance of views with duplicates}.
\newblock \bibinfo{journal}{\emph{SIGMOD Rec.}} \bibinfo{volume}{24}, \bibinfo{number}{2} (\bibinfo{date}{May} \bibinfo{year}{1995}), \bibinfo{pages}{328–339}.
\newblock
\showISSN{0163-5808}
\urldef\tempurl%
\url{https://doi.org/10.1145/568271.223849}
\showDOI{\tempurl}


\bibitem[Gupta and Mumick(1999)]%
        {gupta99maintenance}
\bibfield{author}{\bibinfo{person}{Ashish Gupta} {and} \bibinfo{person}{Inderpal Mumick}.} \bibinfo{year}{1999}\natexlab{}.
\newblock \showarticletitle{Maintenance of Materialized Views: Problems, Techniques, and Applications}.
\newblock \bibinfo{journal}{\emph{Data Engineering Bulletin}}  \bibinfo{volume}{18} (\bibinfo{date}{11} \bibinfo{year}{1999}).
\newblock


\bibitem[Kreps et~al\mbox{.}(2011)]%
        {kreps2011kafka}
\bibfield{author}{\bibinfo{person}{Jay Kreps}, \bibinfo{person}{Neha Narkhede}, \bibinfo{person}{Jun Rao}, {et~al\mbox{.}}} \bibinfo{year}{2011}\natexlab{}.
\newblock \showarticletitle{Kafka: A distributed messaging system for log processing}. In \bibinfo{booktitle}{\emph{Proceedings of the NetDB}}, Vol.~\bibinfo{volume}{11}. Athens, Greece, \bibinfo{pages}{1--7}.
\newblock


\bibitem[Kulkarni et~al\mbox{.}(2014)]%
        {hlc}
\bibfield{author}{\bibinfo{person}{Sandeep~S. Kulkarni}, \bibinfo{person}{Murat Demirbas}, \bibinfo{person}{Deepak Madappa}, \bibinfo{person}{Bharadwaj Avva}, {and} \bibinfo{person}{Marcelo Leone}.} \bibinfo{year}{2014}\natexlab{}.
\newblock \showarticletitle{Logical Physical Clocks}. In \bibinfo{booktitle}{\emph{Principles of Distributed Systems}}, \bibfield{editor}{\bibinfo{person}{Marcos~K. Aguilera}, \bibinfo{person}{Leonardo Querzoni}, {and} \bibinfo{person}{Marc Shapiro}} (Eds.). \bibinfo{publisher}{Springer International Publishing}, \bibinfo{address}{Cham}, \bibinfo{pages}{17--32}.
\newblock
\showISBNx{978-3-319-14472-6}


\bibitem[Leis et~al\mbox{.}(2015)]%
        {leis}
\bibfield{author}{\bibinfo{person}{Viktor Leis}, \bibinfo{person}{Kan Kundhikanjana}, \bibinfo{person}{Alfons Kemper}, {and} \bibinfo{person}{Thomas Neumann}.} \bibinfo{year}{2015}\natexlab{}.
\newblock \showarticletitle{Efficient processing of window functions in analytical SQL queries}.
\newblock \bibinfo{journal}{\emph{Proc. VLDB Endow.}} \bibinfo{volume}{8}, \bibinfo{number}{10} (\bibinfo{date}{June} \bibinfo{year}{2015}), \bibinfo{pages}{1058–1069}.
\newblock
\showISSN{2150-8097}
\urldef\tempurl%
\url{https://doi.org/10.14778/2794367.2794375}
\showDOI{\tempurl}


\bibitem[McSherry(2022)]%
        {mcsherry2022materialize}
\bibfield{author}{\bibinfo{person}{Frank McSherry}.} \bibinfo{year}{2022}\natexlab{}.
\newblock \showarticletitle{Materialize: a platform for building scalable event based systems}. In \bibinfo{booktitle}{\emph{Proceedings of the 16th ACM International Conference on Distributed and Event-Based Systems}}. \bibinfo{pages}{3--3}.
\newblock


\bibitem[McSherry et~al\mbox{.}(2013)]%
        {mcsherry2013differential}
\bibfield{author}{\bibinfo{person}{Frank McSherry}, \bibinfo{person}{Derek~Gordon Murray}, \bibinfo{person}{Rebecca Isaacs}, {and} \bibinfo{person}{Michael Isard}.} \bibinfo{year}{2013}\natexlab{}.
\newblock \showarticletitle{Differential dataflow.}. In \bibinfo{booktitle}{\emph{CIDR}}.
\newblock


\bibitem[Microsoft({[n.\,d.]})]%
        {site-microsoft-mv}
\bibfield{author}{\bibinfo{person}{Inc. Microsoft}.} \bibinfo{year}{[n.\,d.]}\natexlab{}.
\newblock \bibinfo{title}{Azure Synapse Materialized Views}.
\newblock \bibinfo{howpublished}{\url{https://learn.microsoft.com/en-us/sql/t-sql/statements/create-materialized-view-as-select-transact-sql}}.
\newblock
\newblock
\shownote{Accessed: 2024-12-02}.


\bibitem[Shmueli and Itai(1984)]%
        {shmueli84maintenance}
\bibfield{author}{\bibinfo{person}{Oded Shmueli} {and} \bibinfo{person}{Alon Itai}.} \bibinfo{year}{1984}\natexlab{}.
\newblock \showarticletitle{Maintenance of Views.}
\newblock \bibinfo{journal}{\emph{Sigmod Record}}  \bibinfo{volume}{14}, \bibinfo{pages}{240--255}.
\newblock
\urldef\tempurl%
\url{https://doi.org/10.1145/602259.602293}
\showDOI{\tempurl}


\bibitem[Snowflake({[n.\,d.]})]%
        {snowpark}
\bibfield{author}{\bibinfo{person}{Inc. Snowflake}.} \bibinfo{year}{[n.\,d.]}\natexlab{}.
\newblock \bibinfo{title}{Snowpark}.
\newblock \bibinfo{howpublished}{\url{https://www.snowflake.com/en/data-cloud/snowpark/}}.
\newblock
\newblock
\shownote{Accessed: 2024-12-02}.


\bibitem[Spitzer et~al\mbox{.}({[n.\,d.]})]%
        {iceberg-row-lineage}
\bibfield{author}{\bibinfo{person}{Russell Spitzer}, \bibinfo{person}{Nileema Shingte}, {and} \bibinfo{person}{Attila-Peter Toth}.} \bibinfo{year}{[n.\,d.]}\natexlab{}.
\newblock \bibinfo{title}{Spec: Adds Row Lineage}.
\newblock \bibinfo{howpublished}{\url{https://github.com/apache/iceberg/pull/11130}}.
\newblock
\newblock
\shownote{Accessed: 2025-03-25}.


\bibitem[Toshniwal et~al\mbox{.}(2014)]%
        {toshniwal2014storm}
\bibfield{author}{\bibinfo{person}{Ankit Toshniwal}, \bibinfo{person}{Siddarth Taneja}, \bibinfo{person}{Amit Shukla}, \bibinfo{person}{Karthik Ramasamy}, \bibinfo{person}{Jignesh~M Patel}, \bibinfo{person}{Sanjeev Kulkarni}, \bibinfo{person}{Jason Jackson}, \bibinfo{person}{Krishna Gade}, \bibinfo{person}{Maosong Fu}, \bibinfo{person}{Jake Donham}, {et~al\mbox{.}}} \bibinfo{year}{2014}\natexlab{}.
\newblock \showarticletitle{Storm@ twitter}. In \bibinfo{booktitle}{\emph{Proceedings of the 2014 ACM SIGMOD international conference on Management of data}}. \bibinfo{pages}{147--156}.
\newblock


\bibitem[Xu({[n.\,d.]})]%
        {iceberg-variant}
\bibfield{author}{\bibinfo{person}{Aihua Xu}.} \bibinfo{year}{[n.\,d.]}\natexlab{}.
\newblock \bibinfo{title}{Variant Data Type Support}.
\newblock \bibinfo{howpublished}{\url{https://github.com/apache/iceberg/issues/10392}}.
\newblock
\newblock
\shownote{Accessed: 2025-03-25}.


\bibitem[Yan et~al\mbox{.}(2018)]%
        {snowtrail}
\bibfield{author}{\bibinfo{person}{Jiaqi Yan}, \bibinfo{person}{Qiuye Jin}, \bibinfo{person}{Shrainik Jain}, \bibinfo{person}{Stratis~D. Viglas}, {and} \bibinfo{person}{Allison Lee}.} \bibinfo{year}{2018}\natexlab{}.
\newblock \showarticletitle{Snowtrail: Testing with Production Queries on a Cloud Database}. In \bibinfo{booktitle}{\emph{Proceedings of the Workshop on Testing Database Systems}} (Houston, TX, USA) \emph{(\bibinfo{series}{DBTest '18})}. \bibinfo{publisher}{Association for Computing Machinery}, \bibinfo{address}{New York, NY, USA}, Article \bibinfo{articleno}{4}, \bibinfo{numpages}{6}~pages.
\newblock
\showISBNx{9781450358262}
\urldef\tempurl%
\url{https://doi.org/10.1145/3209950.3209958}
\showDOI{\tempurl}


\bibitem[Zaharia et~al\mbox{.}(2012)]%
        {microbatching}
\bibfield{author}{\bibinfo{person}{Matei Zaharia}, \bibinfo{person}{Tathagata Das}, \bibinfo{person}{Haoyuan Li}, \bibinfo{person}{Scott Shenker}, {and} \bibinfo{person}{Ion Stoica}.} \bibinfo{year}{2012}\natexlab{}.
\newblock \showarticletitle{Discretized streams: an efficient and fault-tolerant model for stream processing on large clusters}. In \bibinfo{booktitle}{\emph{Proceedings of the 4th USENIX Conference on Hot Topics in Cloud Ccomputing}} (Boston, MA) \emph{(\bibinfo{series}{HotCloud'12})}. \bibinfo{publisher}{USENIX Association}, \bibinfo{address}{USA}, \bibinfo{pages}{10}.
\newblock


\bibitem[Zhou et~al\mbox{.}(2007)]%
        {Zhou2007LazyMO}
\bibfield{author}{\bibinfo{person}{Jingren Zhou}, \bibinfo{person}{Per-{\AA}ke Larson}, {and} \bibinfo{person}{Hicham~G. Elmongui}.} \bibinfo{year}{2007}\natexlab{}.
\newblock \showarticletitle{Lazy Maintenance of Materialized Views}. In \bibinfo{booktitle}{\emph{Very Large Data Bases Conference}}.
\newblock
\urldef\tempurl%
\url{https://api.semanticscholar.org/CorpusID:971610}
\showURL{%
\tempurl}


\end{thebibliography}
